\documentclass[fleqn,usenatbib]{mnras}

\usepackage{newtxtext,newtxmath}

\usepackage[T1]{fontenc}

\DeclareRobustCommand{\VAN}[3]{#2}
\let\VANthebibliography\thebibliography
\def\thebibliography{\DeclareRobustCommand{\VAN}[3]{##3}\VANthebibliography}

\renewcommand{\citealt}[1]{\citet{#1}}

\usepackage{graphicx}	
\usepackage{amsmath}	
\usepackage{enumitem}
\usepackage{xcolor}
\usepackage{xspace}
\usepackage{braket}

\newcommand{\simba}{\textsc{Simba}\xspace}
\newcommand{\simbaeor}{\textsc{Simba-EoR}\xspace}
\newcommand{\logten}{\log_{10}}

\title[\simbaeor]{\simbaeor: Early galaxy formation in the \simba simulation including a new sub-grid interstellar medium model}

\author[E. Jones et al.]{
E. Jones,$^{1}$\thanks{E-mail: ejones@roe.ac.uk}
B. Smith,$^{1}$
R. Davé,$^{1,2,3}$
D. Narayanan \&$^{4,5}$
Q. Li$^{6}$
\\
$^{1}$Institute for Astronomy, Royal Observatory, University of Edinburgh, Edinburgh EH9 3HJ, UK \\
$^{2}$University of the Western Cape, Bellville, Cape Town 7535, South Africa \\
$^{3}$South African Astronomical Observatories, Observatory, Cape Town 7925, South Africa \\
$^{4}$Department of Astronomy, University of Florida, 211 Bryant Space Sciences Center, Gainesville, FL 32611, USA \\
$^{5}$Cosmic Dawn Center (DAWN), Niels Bohr Institute, University of Copenhagen, Jagtvej 128, København N, DK-2200, Denmark \\
$^{6}$Max-Planck Institute for Astrophysics, Karl-Schwarzschild-Str. 1, D-85741 Garching, Germany
}

\date{Accepted XXX. Received XXX; in original form XXX}

\pubyear{2024}

\begin{document}
\label{firstpage}
\pagerange{\pageref{firstpage}--\pageref{lastpage}}
\maketitle

\begin{abstract}


We update the dust model present within the \simba galaxy simulations with a self-consistent framework for the co-evolution of dust and molecular hydrogen populations in the interstellar medium, and use this to explore $z \geq 6$ galaxy evolution. In addition to tracking the evolution of dust and molecular hydrogen abundances, our model fully integrates these species into the \simba simulation, explicitly modelling their impact on physical processes such as star formation and cooling through the inclusion of a novel two-phase sub-grid model for interstellar gas. Running two cosmological simulations down to $z \sim 6$ we find that our \simbaeor model displays a generally tighter concordance with observational data than fiducial \simba. Additionally we observe that our \simbaeor models increase star formation activity at early epochs, producing larger dust-to-gas ratios consequently. Finally, we discover a significant population of hot dust at $\sim 100$ K, aligning with contemporaneous observations of high-redshift dusty galaxies, alongside the large $\sim 20$ K population typically identified.

\end{abstract}



\section{Introduction}
\label{sec:introduction}

The James Webb Space Telescope (JWST), launched in late 2021, boasts the capability to run astrophysical surveys groundbreaking in their width, depth and accuracy \citep{bib:gardner2006}. Relevant to this study, JWST will observe galaxies at redshifts $z > 6$, collecting state-of-the-art datasets applicable to the research of galaxy formation and evolution. Observable properties taken from JWST images will enable us to quantify parameters critical for a galaxy's evolution, such as its star formation rate (SFR), dust-to-gas ratio, hydrogen abundance and metallicity; giving physical insight into the galactic environment as observed, in addition to its past and future evolution. Furthermore, as the unprecedented depth of JWST observations allows us to probe further into our universe's past with greater accuracy than ever before, it is now possible to observe the evolution of galactic structure during the epoch of reionisation with statistical rigour never-before-seen.

In order to simulate the galaxies that JWST will observe, it is necessary to improve our models of the interstellar medium (ISM) such that they are appropriate for the low-metallicity conditions common to high-redshift galaxies. For example, the \simba simulation has relied upon the Krumholz-McKee-Tumlinson (KMT) model, which becomes inaccurate at metallicities $< 10^{-2} Z_\odot$ and makes assumptions about the dust population that are based on the local Universe \citep{bib:KMT2009, bib:KG2011}. In high-redshift environments it therefore becomes necessary to consider both the creation/evolution of star-forming molecular clouds in the absence of, and during the evolution of, dust populations; as the impact of dust becomes increasingly significant. Contemporary works such as \textsc{thesan} \citep{bib:kannan2022} have made large improvements to the treatment of such high-redshift systems, and in this work we aim to do the same for \simba.

Sub-grid models are necessary to include physical phenomena which occur on scales smaller than the resolution limit of the simulation itself. For example, star formation, which is triggered by the collapse of gas within giant molecular clouds (GMCs) to high densities $\sim 10^6 \,\, \mathrm{cm}^{-3}$ is left unresolved in typical modern cosmological simulations. To solve this, \simba implements a sub-grid model in which the star formation rate (SFR) is proportional to both the gas density and molecular hydrogen fraction. Furthermore, the fraction of molecular hydrogen is also calculated sub-grid, through implementation of the KMT model which assumes the gas is a spherical cloud irradiated uniformly under idealised conditions \citep{bib:KMT2009,bib:krumholz2014}. This model has been fairly successful in producing accurate H$_2$ abundances at higher metallicities, where molecular hydrogen forms primarily on the surface of dust grains, but explicitly only applies down to metallicities of 1\% solar as mentioned previously. Therefore, to model star formation accurately, we must update our treatment of the ISM and model H$_2$ formation explicitly.

Pertinent to this work is the evolution of galactic dust within \simba, modelled using the prescription of \citealt{bib:Li2019}. This is a sub-grid model describing dust growth via the accretion of gaseous metals, and dust destruction via thermal sputtering and supernovae shocks. The grain nuclei are produced (by the condensation of metals ejected during stellar feedback) and annihilated (by astration) externally to the evolution model. Though these processes are implemented within \simba, prior to this work there existed no communication between the dust and molecular hydrogen populations, whose relationship is crucial due to the catalysis of H$_2$ formation on grain surfaces. Given \simba implements a H$_2$-driven star formation scheme, modelling this relationship in a sophisticated manner would boast great improvement to the physicality of the simulation.

As an area of active research, the evolution of galactic dust has seen rapid improvements to its modelling within astrophysical simulations over recent years. A wide range of contemporary studies, varying in model sophistication and numerical application, can provide great insight into the dusty galaxies observed by astronomers. Isolated galaxy simulations \citep{bib:bekki2013, bib:hou2017, bib:aoyama2017, bib:aoyama2018, bib:aoyama2019, bib:mckinnon2018}, for example, are able to implement multi-species dust populations and grain size distributions more accurately than cosmological simulations, owing to their reduced scope and increased resolution on the necessary scales. However, the field is beginning to shift toward implementing these more sophisticated models in full cosmological simulations.

Among the first simulations modelling galaxy-dust co-evolution were the works of \citealt{bib:mckinnon2016, bib:mckinnon2017}, which provided crucial insight regarding the addition of dust to cosmological-scale simulations. Efforts from \citealt{bib:bekki2013, bib:bekki2015, bib:mckinnon2018} sought to model dust processes accurately on galactic scales, going so far as to simulate dust as a `live particle' within the simulation, allowing for the inclusion of gravitational and drag forces. One major issue with treating dust as a `live' particle is the computational overhead introduced by the fact that the gravitational force from the target particle affects all others; the calculation scaling with the number of gravitationally interacting particles in the simulation. In cosmological simulations where the total number of particles is typically very large, such a treatment is not appropriate.

An area of major improvement for dust models in recent years has been the inclusion of a grain size distribution, which tracks the physical size of grains as they grow through accretion and coagulation. For cosmological simulations this is usually achieved through the deployment of a two-size model, where grains are categorised as either large or small, and the population of each is monitored \citep{bib:aoyama2017, bib:hou2016, bib:hou2019, bib:gjergo2018}. A scheme such as this also allows for modelling of coagulation, where grain-grain collisions lead to the formation of larger grains, which many of these works have implemented. The work of \citealt{bib:granato2021} presents an interesting extension of the two-size model with the independent modelling of carbon and silicate dust in cosmological zoom-in simulations. Furthermore, \citealt{bib:aoyama2019} explicitly model 32 different grain radii in their simulation, allowing them to study collisional processes such as accretion growth in much finer detail. In fact, they find that small grains are abundant in the ISM at early times due to their lower accretion efficiency -- this is not something which can be resolved without an explicit grain size distribution. Whilst two-size models have been implemented within cosmological-scale simulations, a full treatment of the grain-size distribution is too computationally expensive to implement.

We have also seen advances in the chemical and thermodynamic treatment of dust within simulations. The works of \citealt{bib:choban2022, bib:choban2024}, for example, present cosmological zooms tracking the evolution of multiple dust species independently. An incredibly sophisticated chemical treatment of dust is presented in \citealt{bib:chiaki2019}, where they model explicitly a 15-species chemical network alongside eight grain compositions which include fully realised size distributions, facilitating the calculation of continuum opacities and gas cooling from radiative grain emission. However, this model was used only to model gas evolution and did not affect star formation. Furthermore, the computational demands of a high-complexity model such as this does not permit its use for large-scale cosmological simulations.

The aim of our work then, is to marry the modelling of star formation, molecular hydrogen and galactic dust in a self-consistent network through which they are inter-dependent as motivated from theory; specifically for use in cosmological simulations. We achieve this by first increasing the capacity of our chemical network such that it tracks explicitly the formation of molecular hydrogen both in the presence and absence of dust. Next, we remove the KMT model, replacing it with the chemically-motivated H$_2$ fraction and introducing a new sub-grid model for the interstellar medium (ISM)\footnote{It is important to note that the KMT model approximates the strength of dust shielding via the metal surface density. We know that this is not entirely accurate because in reality the dust-to-metal ratio is not constant, which warrants an improved treatment of dust shielding. However, Grackle does not currently possess the functionality to compute dust shielding, and as such this mechanism is omitted in our study. Possible avenues for future work on this are to include an attenuation to the ISRF based on the amount of dust present, or implement a full radiative transfer approach.}. Furthermore, we introduce a calculation of the local interstellar radiation field (ISRF) incident on each particle, which is used to compute the dust temperature. Lastly, we update the dust model such that the accretion rate is dependent upon the grain temperature. This affects the dust, and by extension, molecular hydrogen, populations.

In Section \ref{sec:methods} we provide technical descriptions of the calculations used in our models, providing physical explanation where appropriate. Furthermore, we also detail the practical facets of our simulation including the procedures employed in testing and verification. Section \ref{sec:results_fiducialSimba} offers comparison between our \simbaeor simulations, those of fiducial \simba, and practical observations. In Section \ref{sec:results_kmtComparison} we carry out a post-processing procedure on population of galactic particles to quantify the difference between our explicit calculations of the molecular hydrogen fraction ($f_{\mathrm{H}_2}$) using our ISM model, and the KMT-estimated values. Our final set of results are presented in Section \ref{sec:results_dustAndISM} where we investigate the thermodynamic state of the dust and cool-phase ISM gas present in our simulations. Finally, Section \ref{sec:conclusions} provides a summary of our findings alongside proposals for future work.


\section{Methods}
\label{sec:methods}

We run the \simba \citep{bib:simba2019} simulation with updated, self-consistent modelling for dust, molecular hydrogen and star formation; achieved through implementation of Grackle's \citep{bib:grackle2017}\footnote{A modification of Grackle version 3.2.dev2 (found at \href{https://github.com/EwanBJones98/grackle/tree/kiara}{https://github.com/EwanBJones98/grackle/tree/kiara}) was used for this work.} nine-species chemical network and dust thermodynamics. We present two simulations in this work, both are periodic boxes containing $1024^3$ gas and dark matter (DM) particles, but differ in their size. Our larger box has a side-length of 50 Mpc/h, resulting in gas particles of mass $M_\mathrm{gas} = 2.28\times10^6\,\,M_\odot$ and dark matter particles of mass $M_\mathrm{DM} = 1.20\times10^7\,\,M_\odot$. On the other hand, the smaller box has a side length of 25 Mpc/h, which gives $M_\mathrm{gas} = 2.85\times10^5\,\,M_\odot$ and $M_\mathrm{DM} = 1.50\times10^6\,\,M_\odot$.

Both boxes were evolved from $z\sim249-6$ and are cosmologically identical with the following parameters: $\Omega_{m}=0.3$, $\Omega_\Lambda=0.7$, $\Omega_b=0.048$, $H_0=68$~km~s$^{-1}$~Mpc$^{-1}$, and $\sigma_8=0.82$. Furthermore, both simulations used a gravitational softening length of 0.5 Kpc/h, and a friends-of-friends (FOF) linking length of 20\% the mean inter-particle separation for their on-the-fly halo finding. The kernel smoothing length is adaptive such that each particle has 64 neighbouring particles; in doing this we allow the gas kernel to become arbitrarily small, ie. a minimum smoothing length of zero.

The UV background model employed uses the rates of \citealt{bib:haardt2012}, where the values have been recomputed to account for self-shielding. Grackle's self-shielding considers neutral atomic hydrogen and helium, but ignores He$^+$ ionisation and UV background heating -- this behaviour corresponds to \texttt{self\_shielding\_method=3} in the code \citep{bib:grackle2017, bib:rahmati2013}. The table named `HM2021\_shielded' within Grackle holds the shielded UV background values used in this work.

The galaxies and halos presented in our analysis were identified through post-processing of the simulation output using a 6D-FOF search with a linking length of 20\% the mean inter-particle separation. Galaxies are defined to contain at minimum 16 star particles and halos are defined to contain at minimum 24 dark matter particles. For specifics on which packages were used please see the software section at the end of the paper.

    \subsection{Star Formation in \simba}
    \label{sec:methods:starFormation}
    
    \simba implements a stochastic, H$_2$-driven star formation procedure \citep{bib:mufasa2016, bib:simba2019} where, prior to the integration of Grackle's molecular network, the abundance of molecular hydrogen was described by Krumholz, McKee \& Tumlinson's model -- henceforth referred to as the KMT model \citep{bib:KMT2009, bib:KG2011}. The major advantage of this prescription is its low computational cost and simple, two-parameter dependence on the column density and metallicity; both of which are easily calculated during run-time. Whilst this is an appropriate model to use in such simulations, its degree of straightforwardness requires the pretext of many assumptions regarding the gaseous cloud and its environment: we remove such assumptions by explicitly tracking the molecular hydrogen abundance using Grackle \citep{bib:grackle2017} as explained in Section \ref{sec:methods:molecularHydrogen}.
    
    The star formation rate for a gas element is given by,
    \begin{align}
        \frac{d M_\star}{dt} = \epsilon_\star \frac{\rho f_{\text{H}_2}}{t_\text{dyn}}
    \end{align}
    where $\epsilon_\star = 0.02$ is the star formation efficiency, $\rho$ is the mass density, $f_{\text{H}_2}$ is the molecular hydrogen fraction of the particle, and $t_\text{dyn}$ is the dynamical time (the time taken for a homogeneous, spherical gas cloud to collapse in free-fall). Stars are formed in a probabilistic fashion, the likelihood of which primarily depends upon the gas particle's SFR, duration of the current time-step and total gas mass present.

    \subsection{Two-Phase ISM Model}
    \label{sec:methods:twoPhaseISM}

    In order to update the star formation model (see Section \ref{sec:methods:starFormation}), we replaced the KMT model's estimates with the molecular hydrogen fraction as calculated from the chemical network directly. However, as the densities required for star formation are beyond the resolution limits of our cosmological simulations, we introduced a two-phase, sub-grid model for gas in the interstellar medium wherein we model sufficiently dense clouds as being composed of both a warm (low-density) and cool (high-density) phase \citep{bib:springel2003}.

    By default, gas particles represent a population of hot gas at relatively low density. However, once the gas is thought to be capable of forming stars ($n_\mathrm{H} > 0.2 \,\,\text{cm}^{-3}$ in our case) we activate an additional phase within the particle, that of the cool, dense gas from which stars are formed. In practice, this activation is carried out through tracking an additional density and energy field, and assuming that each phase contains half the total mass of the original particle. We allow ISM particles (by which we refer to those with both phases active) to revert to `standard' gas particles if they are no longer potentially star-forming. Each simulation timestep, we update the cool phase density such that it maintains pressure balance with the warm phase, whose properties are in turn updated by the hydrodynamical solver. This pressure balance is expressed mathematically as,
    \begin{align}
        \rho_\text{cool}U_\text{cool} = \rho_\text{warm}U_\text{warm}
    \end{align}
    where $U$ is the internal energy. It is important to note that in the two-phase regime, only the cool phase is explicitly cooled due to its relevance for star formation. The warm phase, the temperature of which is held at $10^4$ K from the equation of state, is used for all other aspects of the simulation such as the hydrodynamic fluid solver. This approach enhances the density of the ISM sufficiently to form stars at rates comparable with the KMT model we replaced, whilst maintaining the self-consistency and accuracy of an explicit chemical network.

    It should be noted that as we model star formation stochastically, (see Section \ref{sec:methods:starFormation}) it is possible that a given gas particle will cool and collapse far beyond the threshold density for star formation, potentially reaching much larger densities before forming a star particle.

    \subsection{Dust Temperature}
    \label{sec:methods:dustTemperature}
    
    The significance of dust in galaxy formation and evolution is not to be understated. The presence of dust grains in the galactic environment acts as a catalyst for molecular formation, namely H$_2$, which is critical for cooling and star formation. Additionally, grains contribute to the thermodynamic state of the system whether that be through: heat exchange with the gas; heating from incident radiation fields, or cooling via thermal emission.

        \subsubsection{Grackle's Solver}
        \label{sec:methods:dustTemperature:grackleSolver}
        
        We employ Grackle \citep{bib:grackle2017} to handle the calculation of the dust temperature; accomplished by using Newton's method to solve the heat balance equation,
        \begin{align}
        \label{eq:heatBalance}
            4\sigma T_{\text{gr}}^4\kappa_{\text{gr}} = 4\sigma T_{\text{rad}}^4\kappa_{\text{gr}} + \Lambda_{\text{gas-grain}}
        \end{align}
        where $\sigma$ is the Stefan-Boltzmann constant, $T_{\text{gr}}$ and $T_{\text{rad}}$ are the temperatures of the dust grains and incident radiation field respectively, $\kappa_{\text{gr}}$ is the grain opacity, and $\Lambda_{\text{gas-grain}}$ is the rate of heat transfer between the gas and dust per unit grain mass. The left-hand and right-hand sides of Equation \ref{eq:heatBalance} describe the cooling and heating contributions respectively. As discussed in \citet{bib:hollenbach1989}, the gas-grain term is a function of both the gas and grain temperatures. Specifically,
        \begin{align}
            \Lambda_{\text{gas-grain}} \propto \left(T_\text{g} - T_{\text{gr}}\right)
        \end{align}
        where $T_\text{g}$ is the gas temperature. From this definition, it can be deduced that when $\Lambda_{\text{gas-grain}} > 0$, the dust is being heated by the gas, hence its appearance on the right-hand side in Equation \ref{eq:heatBalance}.

        The opacity of a grain is modelled as a function of its temperature, as presented in \citealt{bib:dopcke2011},

        \begin{align}
            \kappa_\mathrm{gr} &\propto \begin{cases}
                                T_\mathrm{gr}^2; & T_\mathrm{gr} < 200 \,\, \mathrm{K} \\
                                \mathrm{constant}; & 200 \,\,\mathrm{K} < T_\mathrm{gr} < 1500 \,\, \mathrm{K} \\
                                T_\mathrm{gr}^{-12}; & T_\mathrm{gr} > 1500 \,\, \mathrm{K}
                             \end{cases} \label{eq:dustOpacity}
        \end{align}

    where we normalise $\kappa_\mathrm{gr} = 16 \,\,\mathrm{cm}^2\,\mathrm{g}^{-1}$ at $T_\mathrm{gr} = 200 \,\, \mathrm{K}$ \citep{bib:grackle2017}. The constant regime at intermediate temperatures models the evaporation of the grain's icy mantle, whilst the steep power law at extreme temperatures accounts for dust sublimation \citep{bib:dopcke2011}.

        \subsubsection{Inclusion of an External Radiation Field}
        \label{sec:methods:dustTemperature:externalField}
        
        By default, $T_{\text{rad}}$ in Equation \ref{eq:heatBalance} consists solely of contributions from Cosmic Microwave Background (CMB) radiation. As discussed in Section \ref{sec:methods:modellingISRF}, the ISRF is computed at simulation run-time for each particle from characteristics of its local galactic environment. Therefore, $T_{\text{rad}}$ should incorporate this to improve the calculation of $T_{\text{gr}}$.
            
        Preexisting functionality within Grackle allows us to provide an ISRF by including an additional term in Equation \ref{eq:heatBalance} such that it becomes,
        \begin{align}
        \label{eq:heatBalanceISRF}
            4\sigma T_{\text{gr}}^4\kappa_{\text{gr}} = 4\sigma T_{\text{rad}}^4\kappa_{\text{gr}} + \Lambda_{\text{gas-grain}} + \Gamma_{\text{ISRF}}G_0
        \end{align}
        where $G_0$ is the ISRF field strength in Habing units and $\Gamma_{\text{ISRF}}$ is the dust heating rate resultant from the ISRF \citep{bib:habing1968, bib:krumholz2014}. The addition of this term circumvents the need to modify $T_{\text{rad}}$ directly, and allows us to pass our calculated ISRF to Grackle easily.

    \subsection{Modelling the ISRF}
    \label{sec:methods:modellingISRF}

    Local stellar radiation is the primary source of heating for dust grains in the ISM. Dust is heated by starlight over a wide range of the electromagnetic spectrum, from the UV down into visual wavelengths. When the gas is cooler than the dust, heat is transferred to the gas through the gas-grain collisions discussed above. In our case, to compute the dust temperature, the local ISRF strength is a necessary input and therefore must be calculated at run-time. This is a non-trivial exercise, as the variables required for its direct computation (e.g. stellar luminosity) are not tracked during the simulation.

    Our work implements an on-the-fly scheme to calculate the ISRF incident upon a given gas particle; allowing us to account for the state of the local stellar environment at the current time-step. We model the ISRF as being proportional to the specific star formation rate (sSFR), where the proportionality is assumed to be that of the Milky-Way,
    \begin{align}
        \label{eq:localISRF}
        G_{0,i} = \frac{G_{0,\text{MW}}}{\mathrm{sSFR}_\mathrm{MW}}\mathrm{sSFR}_{i}
    \end{align}
    where $i$ and MW denote the target particle and Milky-Way respectively and the total sSFR within the target particle's local environment is denoted sSFR$_i$. Specifically, we use $\mathrm{sSFR}_\mathrm{MW} = 2.71\times10^{-11} \,\, \mathrm{yr}^{-1}$ \citep{bib:licquia2015} and $G_{0,\mathrm{MW}} = 0.6 \,\, \mathrm{Habing}$ \citep{bib:sigame2017}. For reference: $1 \,\, \mathrm{Habing} \equiv 1.6\times10^{-3}\,\,\mathrm{erg}\,\mathrm{s}^{-1}\,\mathrm{cm}^{-2}$.
    
    To calculate the specific star formation rate for a given particle we first find the total star formation rate and stellar mass within its kernel. Dividing these quantities gives the sSFR of the particle. Once every particle has performed this computation, we proceed to calculate the final value for each particle as the kernel-weighted average of all sSFR's within a smoothing length. This is an attempt to minimise the impact of individual star particles with extremely high star formation rates.
    
    In the future, it would be ideal to use a more sophisticated approach as the ISRF is largely responsible for dust heating. One method would be to use an external code such as \texttt{STARBURST99} \citep{bib:STARBURST99} which is able to generate estimates of the luminosity for a range of input parameters; notably the initial mass function (IMF), stellar age and metallicity. By interpolating over a grid of suitable parameter values, a lookup table for the luminosity can be created and later referenced during simulation run-time. Calculation of the ISRF using this estimate is then straight forward, for example, employed in \citet{bib:sigame2017} is,
    \begin{align}
        G_{0} = \sum_{\rvert \Vec{r}_{\text{gas}} - \Vec{r}_{\star, i} \rvert < h} \frac{L_{\star}}{4\pi \rvert \Vec{r}_{\text{gas}} - \Vec{r}_{\star, i} \rvert^2}\frac{M_{\star}}{10^4M_{\odot}}
    \end{align}
    where $L_{\star}$ is the stellar luminosity, $\rvert \Vec{r}_{\text{gas}} - \Vec{r}_{\star, i} \rvert$ is the separation between the target gas particle and stellar source, $M_{\star}$ is the stellar mass of the source. Implementation of such a calculation is, however, challenging given the resolution constraints of a cosmological simulation -- hence its non-inclusion in this work.
       
    \subsection{Dust Growth and Destruction}
    \label{sec:methods:dustGrowthAndDestruction}

    Having implemented models for both the incident ISRF and dust temperature of a given particle, the final step is to include the modelling of dust. Dust is produced in our simulation identically to fiducial \simba, with grains forming through the condensation of metals ejected from supernovae and AGB (asymptotic giant branch) stars into the ISM \citep{bib:simba2019, bib:Dwek1998}.
    
    Once formed, the dust population is evolved through consideration of three processes: growth via metal accretion, destruction via thermal sputtering and destruction via supernovae shocks. To model these processes this we implement the work of \citealt{bib:Li2019}, highlighting any deviations from their method explicitly in the following sections.
    
    Furthermore, as modelled in \simba, grains are completely destroyed via astration \citep{bib:mattsson2021, bib:dayal2022} and in the presence of hot winds \citep{bib:simba2019}. Any further processes, such as grain-grain shattering and non-thermal sputtering, not prescribed in \citealt{bib:Li2019} or \citealt{bib:simba2019} are omitted in this work.
    
        \subsubsection{Metal Accretion}
        \label{sec:methods:dustGrowthAndDestruction:metalAccretion}
    
        Dust grains grow in size through the accretion of gaseous metals, the rate of which is given by,
        \begin{align}
        \label{eq:dustGrowth}
            \left( \frac{dM_{\text{gr}}}{dt} \right)_{\text{growth}} =  \left( 1 - \frac{M_{\text{gr}}}{M_{\text{met}}} \right) \frac{M_{\text{gr}}}{\tau_{\text{accr}}}
        \end{align}
        where $M_{\text{gr}}$ is the dust grain mass, $M_{\text{met}}$ is the total metal mass in both the gaseous and dust phases \citep{bib:Dwek1998}. Furthermore, $\tau_{\text{accr}}$, the accretion timescale takes the form,
        \begin{align}
        \label{eq:dustGrowthTimescale}
            \tau_{\text{accr}} = \tau_{\text{ref}}\left(\frac{T_\text{ref}}{T_\text{g}}\right)^{1/2} \left(\frac{\rho_\text{ref}}{\rho_\text{g}}\right) \left(\frac{Z_\odot}{Z_\text{g}}\right)
        \end{align}
        where $\tau_\text{ref}$, $Z_\odot$ and $\rho_\text{ref}$ are the reference timescale, solar metallicity and density respectively; analogous variables for the gas are subscripted `g' \citep{bib:Li2019,bib:Asano2013,bib:Hirashita2000}\footnote{Please note that the original publication \citep{bib:Li2019} contains a typographic error in that the square root on the temperature term is omitted. The equation presented here corrects this.}. These constants are set to the following values in this work:
        \begin{align*}
            \tau_\text{ref} &= 0.4 \,\,\,\, \text{Myr} \\
            \rho_\text{ref} &= 2.3 \times 10^{-22} \,\,\,\, \text{g}\,\text{cm}^{-3} \\
            Z_\odot &= 0.01295
        \end{align*}
        It may be useful to note that $\rho_\text{ref} \sim 100 \,\, \text{m}_\text{H} \, \text{cm}^{-3}$, and the solar metallicity value originates from Cloudy, version 13 \citep{bib:ferland2013}. Although the choice of these values is somewhat arbitrary, they have been chosen to represent the `common' physical scenario.
        
        There is one key difference in our model versus the \citet{bib:Li2019} model -- we replace the reference temperature ($T_\mathrm{ref}$ = 20 $\mathrm{K}$ originally) with the dust temperature. This is done as a crude representation of the dust temperature's influence on the accretion rate; specifically the sticking potential of metals onto grains \citep{bib:hirashita2011, bib:spitzer2004}. The exact form of the sticking potential is difficult to obtain with its many complex dependencies on the chemical composition of the grains in addition to the environment in which they preside \citep{bib:jones2011}. As our model does not attempt to represent the distribution of grain sizes, nor their chemical composition, integrating an accurate sticking potential is impracticable in this work. Furthermore, we must be careful to retain the characteristics of the original model, altering the accretion rate's temperature dependence too much would constitute a new model itself and is not the aim of this work. We re-tune $\tau_\mathrm{ref}$ such that our model aligns with expected results for the stellar mass function at redshift 6. Exact details of these changes are presented in Appendix \ref{app:dustGrowth}, alongside a derivation of Equation \ref{eq:dustGrowthTimescale}.
        
        It bears clarification that this model assumes a constant radius of $0.1 \,\, \mu\mathrm{m}$ for all dust grains, omitting the inclusion of a grain size distribution. Thus, when dust grains grow/shrink in our model, we are referring to a change in their mass and metallicity; their cross-section (relevant in the formation of molecular hydrogen, see Section \ref{sec:methods:molecularHydrogen:dustCase}) remains constant. Models which account for the distribution of grain sizes are a current area of interest, and such future enhancements will make for an appreciable update to our model. We refer the curious reader to \citealt{bib:romano2022}, where both molecular hydrogen and dust are modelled using a distribution of grain sizes.
        
        \subsubsection{Dust Sputtering}
        \label{sec:methods:dustGrowthAndDestruction:dustSputtering}

        Thermal sputtering\footnote{For a full discussion on dust sputtering, including non-thermal sputtering, please see \citet{bib:Hu2019}} is the process by which dust grains moving within a hot gas are abraded due to the collisions between the dust and gas phases: a result of the gas' high temperature \citep{bib:draine1979, bib:tielens1994}. The sputtering timescale is formally defined as $\tau_\text{sput} = a / \left| \dot{a} \right|$, where $a$ is the grain radius. However, as we do not model a grain size distribution, $\tau_\mathrm{sput}$ is approximated as a function of the gas density and temperature within each gas element: for a comprehensive discussion of this please see \citet{bib:tsai1995} and \citet{bib:Li2019}. With the sputtering timescale calculated, the rate at which dust grains grow due to sputtering is given by,
        \begin{align}
            \left( \frac{dM_\text{gr}}{dt} \right)_\text{sput} = - \frac{M_\text{gr}}{\tau_\text{sput}/3}.
        \end{align}

        \subsubsection{Destruction via Shocks}
        \label{sec:methods:dustGrowthAndDestruction:shockDestruction}
        
        Thermal sputtering of dust (see Section \ref{sec:methods:dustGrowthAndDestruction:dustSputtering}), whilst able to completely destroy small grains through continued erosion, is not sufficient for larger grains. The destruction of these large grains is primarily driven by shocks originating from supernova remnants, SNRs \citep{bib:slavin2015}. The timescale for this destruction is given by,
        \begin{align}
        \label{eq:shockDestruction}
            \tau_\text{de} = \frac{M_\text{g}}{\epsilon \gamma M_\text{s}}
        \end{align}
        where $\epsilon = 0.3$ is the grain destruction efficiency from Type-II Supernovae, SNII, \citep{bib:mckee1989}, $\gamma$ is the rate at which SNII occur in the local environment, and $M_\text{s}$ is the mass of the shocked gas, per supernova, travelling at $100 \,\, \text{km} \, \text{s}^{-1}$ or faster. It is assumed in these definitions that SNI (Type-I supernovae) are negligible relative to SNII due to the particular circumstances under which they occur, and their subsequent low explosion frequency. In practice, the SNII rate is passed to the model from the simulation itself, whilst the mass of shocked gas is calculated using the Sedov-Taylor approximation.

    \subsection{The Abundance of Molecular Hydrogen}
    \label{sec:methods:molecularHydrogen}
    
    The abundance of molecular hydrogen within the ISM is largely dependent upon the amount of dust present. In the absence of dust, molecular hydrogen struggles to form efficiently due to its nucleic symmetry and subsequent lack of dipole transition. Upon the collision of two neutral hydrogen atoms, in the case that they bond to form H$_2$, their excess kinetic energy will cause the molecule to exist in an unstable state. Unless this excess energy is radiated from the molecule, it will dissociate and the constituent atoms will separate. Due to the absence of dipole transition, H$_2$ must radiate via forbidden magnetic quadrupole transition. The lowest-lying transition of this type has a transition probability of $\sim 10^{-11} \,\, \text{s}^{-1}$, which is far too unlikely to occur within the lifetime of the unstable molecule \citep{bib:bromm2013}. 
        
    Nevertheless, there are chemical pathways through which molecular hydrogen can form in the absence of dust (such as in the primordial case), although they are much slower. A brief discussion of both scenarios alongside their implementations will be presented below due to the novel nature of their inclusion within \simba (see Section \ref{sec:methods:starFormation}).

        \subsubsection{The Primordial Case}
        \label{sec:methods:molecularHydrogen:primordialCase}
        In the primordial gas there exist three primary channels through which molecular hydrogen forms \citep{bib:galli1998}; the first of which we will consider being the following,
        \begin{align*}
            &\text{H} + \text{e}^- \longrightarrow \text{H}^- + \gamma\\
            &\text{H}^- + \text{H} \longrightarrow \text{H}_2 + \text{e}^-
        \end{align*}
        wherein a free electron associates with a neutral hydrogen atom, creating H$^-$, which is stable due to its permanent electric dipole and resultant radiative transition. This proceeds form H$_2$ upon collision with another neutral hydrogen atom, which forms a stable molecule due to the ejection of the free electron which carries any excess energy; therefore removing the need for fast radiative emission. As the first reaction above requires collision with a free electron, the ionisation factor of the gas is primary in determining the effectiveness of this pathway.

        At high redshifts, $z \gtrsim 100$, photons from the cosmic microwave background (CMB) are sufficiently energetic to dissociate the weakly-bound H$^-$ species necessary in the formation channel above. However, these photons are unable to destroy H$_2^+$ and so we consider the second pathway,
        \begin{align*}
            &\text{H}^+ + \text{H} \longrightarrow \text{H}_2^+ + \gamma\\
            &\text{H}_2^+ + \text{H} \longrightarrow \text{H}_2 + \text{H}^+
        \end{align*}
        where the ejection of a proton removes any excess energy from the newly formed molecule. This channel is significantly slower than the first and as such is only important in the regime under which the first cannot operate \citep{bib:bromm2013, bib:tegmark1997}.

        The final pathway we will detail here is that of the three-body reaction,
        \begin{align*}
            \text{H} + \text{H} + \text{H} \longrightarrow \text{H}_2 + \text{H}
        \end{align*}
        which only becomes effective at high densities ($n \gtrsim 10^8 \,\, \mathrm{cm}^{-3}$) where the rate at which three-body collisions occur is non-negligible. Furthermore, as the density of the gas increases, so too does the rate of ion-electron recombination -- ultimately reducing the density of free electrons in the gas and therefore the rate of H$_2$ formation in the first channel \citep{bib:turk2011}.

        \subsubsection{In the Presence of Dust}
        \label{sec:methods:molecularHydrogen:dustCase}
        
        Dust grains are the primary channel through which molecular hydrogen is formed within the collapsing gaseous cloud described above. The grains present within a dust-enriched cloud are sites of active H$_2$ formation due to their ability to capture hydrogen atoms on their surface, absorbing the atom's kinetic energy upon collision. Over time, a large population of hydrogen atoms will accumulate on the surface of a grain. Able to move slowly on the grain surface, these atoms will collide with one-another to form H$_2$. Excess energy which threatens to destabilise the molecule will be used to overcome the sticking potential of the grain, allowing the hydrogen molecule to leave the grain surface and re-enter the gas phase.

        The dust-catalysed H$_2$ formation rate, as prescribed in \citet{bib:tielens1985}, can be expressed as,
        \begin{align}
        \label{eq:h2DustRateExact}
            R = \frac{1}{2}\overline{v}_\text{H}n_\text{H}n_\text{gr}\sigma_\text{gr}\epsilon_{\text{H}_2}S(T) \,\,\,\, \text{cm}^{-3}\,\text{s}^{-1}
        \end{align}
        where $\overline{v}_\text{H}$, $n_\text{H}$, $n_\text{gr}$, $\sigma_\text{gr}$, $\epsilon_{\text{H}_2}$ and $S(T)$ are the average velocity and number density of hydrogen atoms, number density of dust grains, grain cross-section, efficiency of grain-bound molecular hydrogen formation and the sticking coefficient of hydrogen atoms respectively \citep{bib:schneider2006}. Due to the complex dependencies present in Equation \ref{eq:h2DustRateExact} we employ the following approximation shown in \citet{bib:hollenbach1989},
        \begin{align}
        \label{eq:h2DustRateApprox}
            R \simeq 6.39\times10^{-15}\left(\frac{T}{300}\right)^{1/2}n_\text{gr}n_\text{H}S(T,T_\mathrm{gr}) \,\,\,\, \text{cm}^{-3} \, \text{s}^{-1}
        \end{align}
            
        The sticking coefficient describes how readily hydrogen atoms will become bound to the grain surface and is thus dependent on both the gas and grain temperatures as detailed in \citet{bib:hollenbach1979},
        \begin{align}
        \label{eq:stickingCoefficient}
            S(T,T_\mathrm{gr}) = \left[ 1 + 0.4\left( \frac{T + T_\text{gr}}{100 \, \text{K}} \right)^{1/2} + 0.2 \left(\frac{T}{100 \, \text{K}}\right) + 0.08 \left(\frac{T}{100 \, \text{K}}\right)^2 \right]^{-1}.
        \end{align}

\section{Comparison with Observation and Fiducial \simba}
\label{sec:results_fiducialSimba}

In this section we offer comparison between observed results, \simba's fiducial modelling prescription, and the new models presented in this work. We include both simulation sizes: the larger box of side-length 50 Mpc/h containing 2 $\times$ 1024$^3$ particles (baryonic and dark matter); and the smaller 25 Mpc/h box with the same number of particles, resulting in a simulation with 8$\times$ higher mass resolution. These runs are labelled 'm50n1024' and 'm25n1024' respectively. All results presented in this section are taken from the redshift 6 snapshot unless explicitly stated otherwise.

    \subsection{Mass Functions}
    \label{sec:results_fiducialSimba:massFunctions}

    \begin{figure*}
        \centering
        \includegraphics[width=1.75\columnwidth]{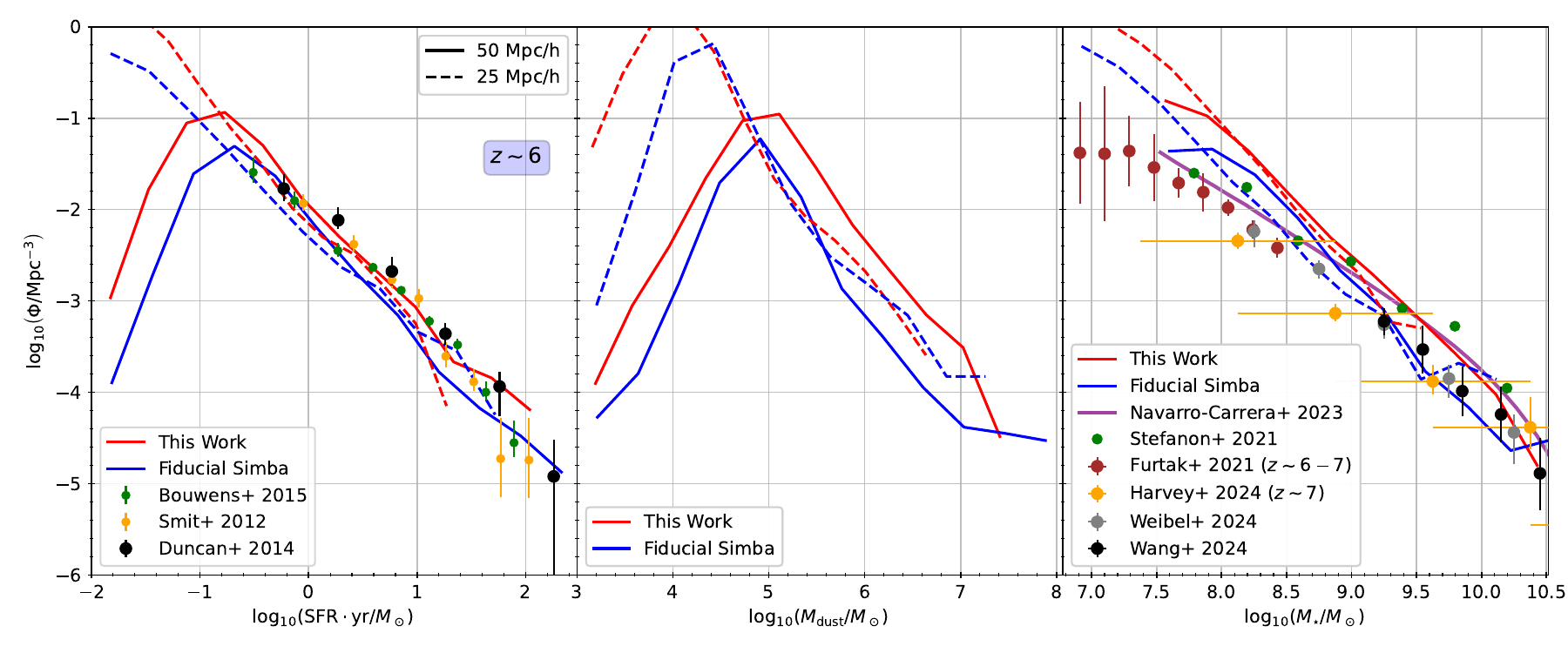}
        \caption{The mass functions for the star formation rate (left), dust mass (centre) and stellar mass (right) at redshift 6. The blue and red lines show results from fiducial \simba and this work's runs respectively. Further, solid/dashed lines differentiate between the 50/25 Mpc/h simulations. \textbf{Left:} We add the observational data presented in \citealt{bib:smit2012}, \citealt{bib:duncan2014} and \citealt{bib:bouwens2015} for comparison with our simulations and see a high level of agreement between all for $\mathrm{SFR} \gtrsim 10^{-1} \,\, M_\odot/\mathrm{yr}.$ At low star formation rates we see considerable divergence with resolution. \textbf{Centre:} We observe similar trends as were seen for the SFR; the box-sizes diverging significantly at $M_\mathrm{dust} \lesssim 10^{4.8} \,\, M_\odot$. It is expected that there is a tight correlation between SFR and dust mass due to its importance in the cooling of star-forming regions. \textbf{Right:} Data from \citealt{bib:stefanon2021}, \citealt{bib:furtak2020}, \citealt{bib:navarroCarrera2023}, \citealt{bib:harvey2024}, \citealt{bib:weibel2024} and \citealt{bib:wang2024} is plotted alongside the simulation results to show their physicality; all of these are at redshift 6 unless specified otherwise in the legend. Tight agreement between the data and simulations is seen at larger stellar masses, whilst the simulations overestimate the data at low stellar mass. For each resolution it can be seen that our models produce more stellar mass than fiducial \simba across the domain.}
        \label{fig:massFunctions}
    \end{figure*}

    In Figure \ref{fig:massFunctions} we plot the SFR (left panel), dust mass (centre panel) and stellar mass (right panel) functions for galaxies identified in both our simulations (red) and the fiducial \simba runs (blue). The size of the simulation box is displayed by solid and dashed lines for the 50 and 25 Mpc/h runs respectively. Where appropriate, observational results have been added to serve as a reference for comparison between datasets. Observing the three panels we identify some general trends: our models predict larger number densities than their fiducial counterparts; the high-resolution boxes contain significantly more data at the lowest values; all four simulations display a high level of agreement throughout the intermediate-high values; and all simulated datasets effectively reproduce observational results.  
    
    The divergence at low values arises purely due to numerical resolution; the higher resolution volume is able to resolve smaller systems.  On the other hand, one can also see that the larger volume tends to have higher number densities at the very highest values.  This is because the smaller volume is less able to produce a representative sample of the most mass massive systems.  Therefore the most robust predictions are in the intermediate galaxy size range, and we will focus our discussion there.

    The leftmost panel describes the distribution of star formation rates produced by the simulations, in addition to various observational results for comparison. Here, it appears that robust predictions arise with the range of $10^{-0.6} \,\, M_\odot/\mathrm{yr} \lesssim \mathrm{SFR}  \lesssim 10^{1.2} \,\, M_\odot/\mathrm{yr}$, which is resolved in both simulations but is not so large as to be subject to stochasticity due to the finite volume.  Within this range, we see the slope and amplitude of the observational SFR functions is broadly well-reproduced by all the simulations.  Also, there is generally good concordance between the predictions of the two volumes, suggesting that the SFR function is reasonably well resolution-converged.
    
    That said, closer inspection reveals that within the robustly predicted range, the \simbaeor simulations (red) yield somewhat better agreement with the observations as compared to the \simba runs (blue). \simba tends to produce a steeper slope than observed, and insufficient high-SFR galaxies.  This is important because \simba has difficulty reproducing the very brightest early galaxies~\citep{bib:finkelstein2023}, and this suggests that \simbaeor may do a bit better; we will explore this further when we compare directly to UV luminosity functions next.  Also, for both models, the 25 Mpc/h runs tends to predict values that are $\sim 0.1$~dex lower than the 50 Mpc/h runs, indicating a slight level of non-convergence even within the well-resolved regime.  
   
    The dust mass function (shown in the centre panel) is seen to have somewhat different trends to the SFR function. Firstly, we identify robust predictions within the range of $10^{4.8} \lesssim M_\mathrm{dust} / M_\odot \lesssim 10^{6.6}$, where all runs appear to give resolved predictions. Here, we see larger discrepancies between the new \simbaeor predictions and \simba, at least for the 50 Mpc/h run. Interestingly, the smaller volume shows substantially less differences; it is not entirely clear why. Neither volume is all that well resolution-converged, but in different senses: \simbaeor's larger volume produces more dust, while \simba's produces less.  Given the complex interplay between the dust model, star formation, metallicity growth, and other physical processes, it is not straightforward to interpret these trends. However, it is clear that all the models produce significant masses of dust in the early Universe. Therefore it will be important to account for dust extinction when examining forward-modelled properties in the observational plane. 
    
    The galactic stellar mass function, shown in the right-hand panel, offers robust predictions in the range $10^{7.8} \lesssim M_\star \lesssim 10^{9.8} \,\, M_\odot$ where the simulations are resolved. We see that the simulated runs, \simba and \simbaeor, exhibit steeper gradients than the observational datasets present. Whilst all runs predict larger populations than observed at low stellar mass, as the stellar mass increases the computational models begin to converge toward the observed populations. At larger stellar masses, $M_\star \sim 10^{9.6} \,\, M_\odot$ we good agreement between the \simbaeor model and both observational datasets. Whilst the fiducial \simba model displays noticeable discrepancies between the 25 Mpc/h and 50 Mpc/h boxes across much of the resolved domain, the \simbaeor boxes show good agreement, highlighting our model's robustness. Furthermore, the \simbaeor model contains more sources at each stellar mass than fiducial \simba, a reflection of its generally higher star formation rate as shown in the leftmost panel.

   In summary, we see general agreement between the results of our model and the fiducial \simba model. Given that the fiducial model produced results which were physically viable itself, agreement with this is an ideal outcome for our model given its reduced parameterisation and self-consistent nature which removed the need for many assumptions previously used. The area where we see signs of our model offering improvement over its predecessor is around the resolution limit; though the impact of this is minimal.

    \subsection{The UV Luminosity Function}
    \label{sec:results_fiducialSimba:UVLuminosityFunction}

    \begin{figure}
        \centering
        \includegraphics[width=\columnwidth]{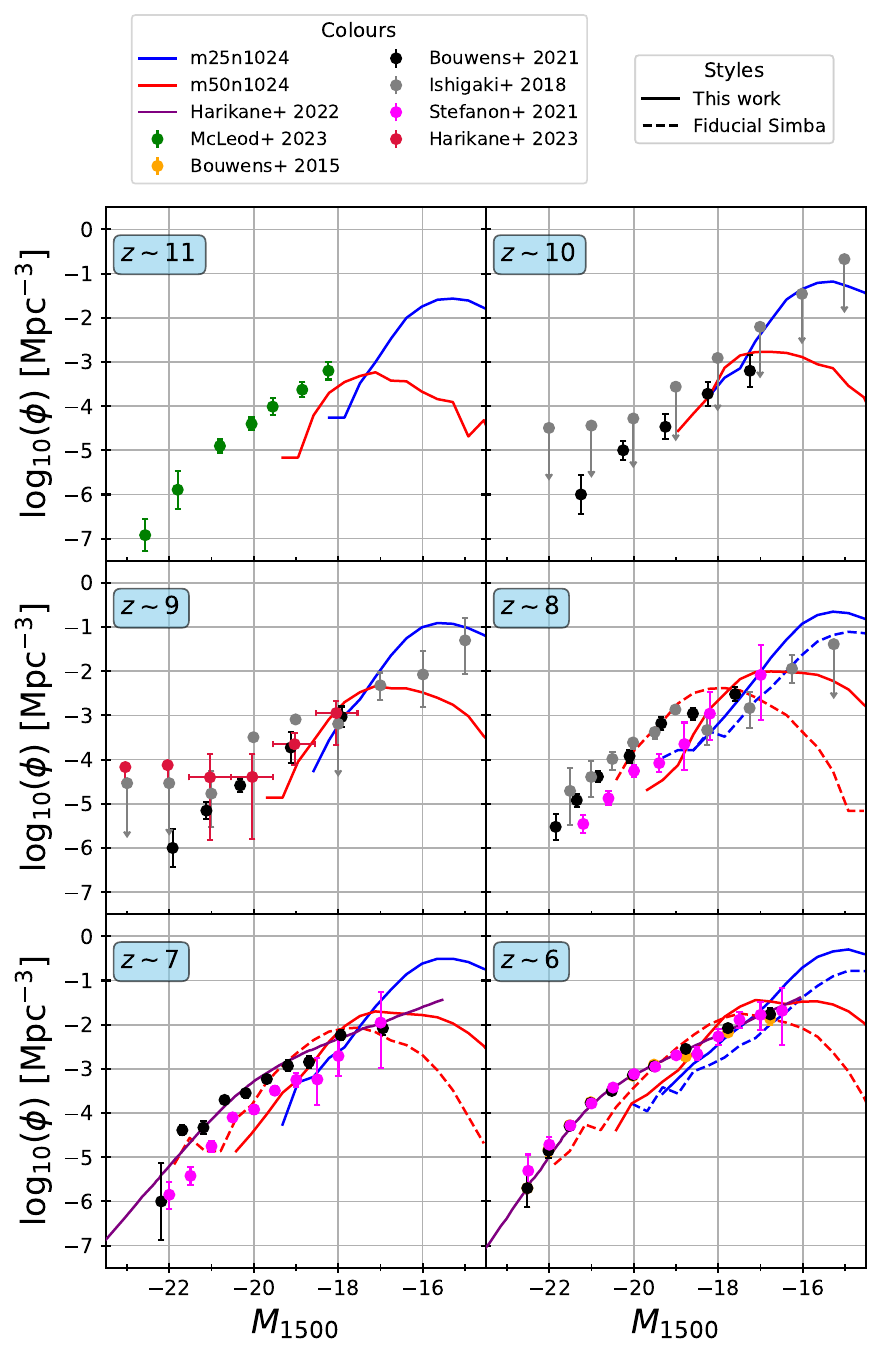}
        \caption{The UV luminosity function at varying redshift for both our 50 Mpc/h (red) and 25 Mpc/h (blue) boxes. Where available, we plot fiducial \simba (dashed lines) predictions alongside our own (solid lines) for comparison. Furthermore, we include observational results from: \citealt{bib:bouwens2015,bib:bouwens2021} as yellow and black points respectively; \citealt{bib:ishigaki2018} as grey points; \citealt{bib:stefanon2021} as pink points; \citealt{bib:harikane2022, bib:harikane2023} as a purple line of best fit and crimson datapoints respectively; and lastly \citealt{bib:mcleod2023} as green datapoints. These points are plotted on the axis corresponding to their redshift. We see that our model offers acceptable predictions, which are in general agreement with the observed data for the population of faint galaxies. Our model is seen to underestimate the number of bright sources, which may be in part due to the increased presence of dust shown in Figure \ref{fig:massFunctions}.}
        \label{fig:UVluminosityFunction}
    \end{figure}

    The UV luminosity functions for the simulations presented in this work are shown in Figure \ref{fig:UVluminosityFunction}. We differentiate between the simulated box-size using red lines for the 50 Mpc/h run and blue lines for 25 Mpc/h run, and between the \simbaeor and fiducial \simba models with solid and dashed lines respectively. Furthermore, we add an array of observational datasets to the panels at their respective redshifts, as described in the figure caption. We plot the luminosity function from redshift $\sim 11$ to $\sim 6$ in integer steps to observe its temporal evolution. It should be noted that we have assumed a composite extinction law to produce the photometric data used in this paper. This uses a Milky-Way extinction law for galaxies with $\mathrm{sSFR} < 0.1\,\,\mathrm{Gyr}^{-1}$ and a Calzetti extinction law for galaxies with $\mathrm{sSFR} > 1\,\,\mathrm{Gyr}^{-1}$ -- intermediate galaxies use a linear combination of the two.

\begin{figure}
        \centering
        \includegraphics[width=0.65\columnwidth]{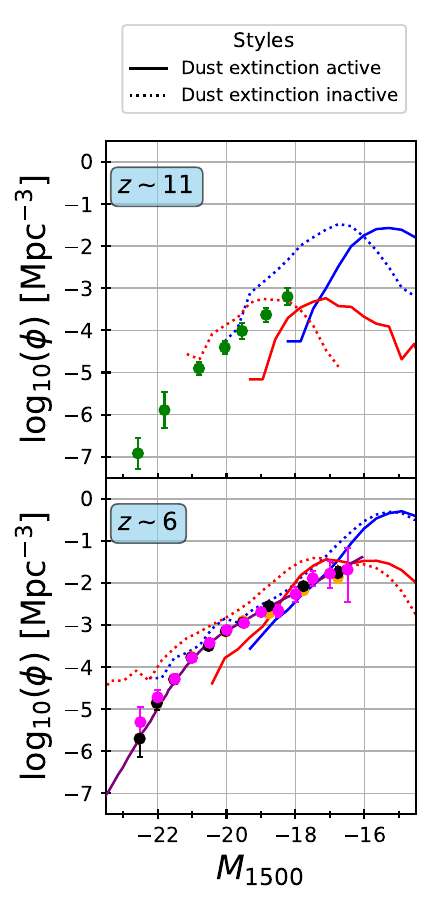}
        \caption{The UV luminosity function at redshift $\sim 11$ and $\sim 6$ for both our simulated boxes. We show the 50 Mpc/h and 25 Mpc/h runs in red and blue respectively, alongside the observational data presented in Figure \ref{fig:UVluminosityFunction}. Furthermore, we display the UV luminosity function as calculated from the absolute magnitude with solid lines, whilst that calculated from the absolute magnitude omitting the dust properties with dashed lines. We see that there exists a significant difference between the dust present/absent luminosity functions at both redshifts, with the dust-obscured samples containing fewer high-brightness sources than their dust-absent counterparts.}
        \label{fig:dust_UVluminosityFunction}
    \end{figure}

    To compute these UV magnitudes, a post-processed ray tracing procedure was ran using \textsc{Pyloser}, which calculates the dust extinction explicitly. This uses the dust properties of each particle as calculated in the simulation, meaning that the models implemented have a direct impact on the UV luminosity function. For example, if dust were ignored entirely, the UV luminosity function would show many more bright sources. On the other hand, if the dust populations were overestimated (if grains grew too quickly for example) we would expect that the UV luminosity function would further underestimate the number of bright sources. We show the affect of dust extinction on the UV luminosity function for our simulations in Figure \ref{fig:dust_UVluminosityFunction}.
    
    First considering only the $z \sim 6$ function, we observe acceptable agreement with the observational data at both simulated resolutions for the dimmest sources, whilst our models are shown to underestimate the population of bright sources, though discrepancies never exceed $\sim 1$ dex. We see that whilst the 50 Mpc/h box contains sources brighter than its counterpart, its population of dim sources is diminished. This is a result of the simulation box-size (as larger boxes contain more massive halos/galaxies than smaller ones) and not a result of our models: as evidenced by the same phenomena occurring in the fiducial runs. Furthermore, turnovers seen in the faint sources for some simulations is a result of numerical resolution in our simulations (akin to those present in the mass functions previously discussed in Section \ref{sec:results_fiducialSimba:massFunctions}), and as such we should only draw results in the resolved region at lower magnitudes than the turnover's peak.

    As the redshift decreases from $\sim 11$ we see that, in general, the disparity between our simulations and observation ameliorates. Focusing on our 50 Mpc/h box, at $z\sim11$ our simulation fails to produce any bright sources, although the population of sources at $M_{1500}\sim-19$ are within $\sim1$ dex of the observations. Generally, we see that our results under-predict the observed data at this redshift. This shortfall is also noted in \citealt{bib:finkelstein2023}, where most models considered show such a deficit, and it is posited that sample contamination may lead to an overabundance of observed sources.

    By redshift 8 we see that our simulations are in good agreement with available datasets for sources of moderate and dim brightness, though these observations do show scatter of $\sim1$ dex between themselves. As the redshift further decreases towards $z \sim 6$ we see that our simulations begin to produce a higher population of bright galaxies, though still underestimating their number as expected from observation. The 25 Mpc/h box shows much the same evolution, however its skew toward dimmer galaxies results in lower populations of moderately bright sources when compared to the larger box. From this analysis we learn that the size of the simulated region is crucial in generating high-brightness galaxies and expect that running our model in a larger box will produce a better fit to observation. Furthermore, observations themselves are limited by the brightness of objects (especially at the high redshifts presented here) and so only the most luminous galaxies will be observed, whilst the large population of dim galaxies are overlooked.

    To understand the effect of dust on the UV luminosity function, we show in Figure \ref{fig:dust_UVluminosityFunction}, the dust-obscured (solid lines) and dust-absent (dashed lines) results at redshifts $\sim 11$ and $\sim 6$. As in Figure \ref{fig:UVluminosityFunction}, the 50 Mpc/h and 25 Mpc/h runs are shown in red and blue respectively, alongside various observed datasets. At both redshifts, we observe an increase in the number density of bright sources when the presence of dust is disregarded: as mentioned above, this is to be expected.

    Crucially, this figure highlights the significance of dust in the luminosity function, with the disparity between the dust-free and dusty cases manifesting directly in the plot. We immediately see that, at both redshifts, the omission of dust extinction raises the number of bright sources substantially, resulting in populations closer to those observed. Interestingly, whilst the $z\sim11$ results show a diminished population of dim galaxies when dust is ignored, the $z\sim6$ results show similar populations of dim galaxies between the two cases. This is a manifestation of the artificial, high-redshift feedback suppression present in \simba. This suppression is achieved by lowering the mass loading factor $\eta$, ie. reducing the mass lost via winds. As the metallicity depends on the mass loading factor, $Z \sim \left(1+\eta\right)^{-1}$, this artificial feedback suppression ultimately increases the metallicity beyond what is typically expected \citep{bib:finlator2008}. In turn, this metal enrichment facilitates early dust growth/formation, leading to greater dust extinction. In short, a large number of dim galaxies at $z\sim11$ are richer with dust than conventional wisdom would suggest, however this is a result of the high-redshift feedback treatment within \simba; a limitation which requires improvement in the future.

    \subsection{The Time Evolution of the Star Formation Rate Density}
    \label{sec:results_fiducialSimba:sfrTimeEvolution}

    \begin{figure}
        \centering
        \includegraphics[width=\columnwidth]{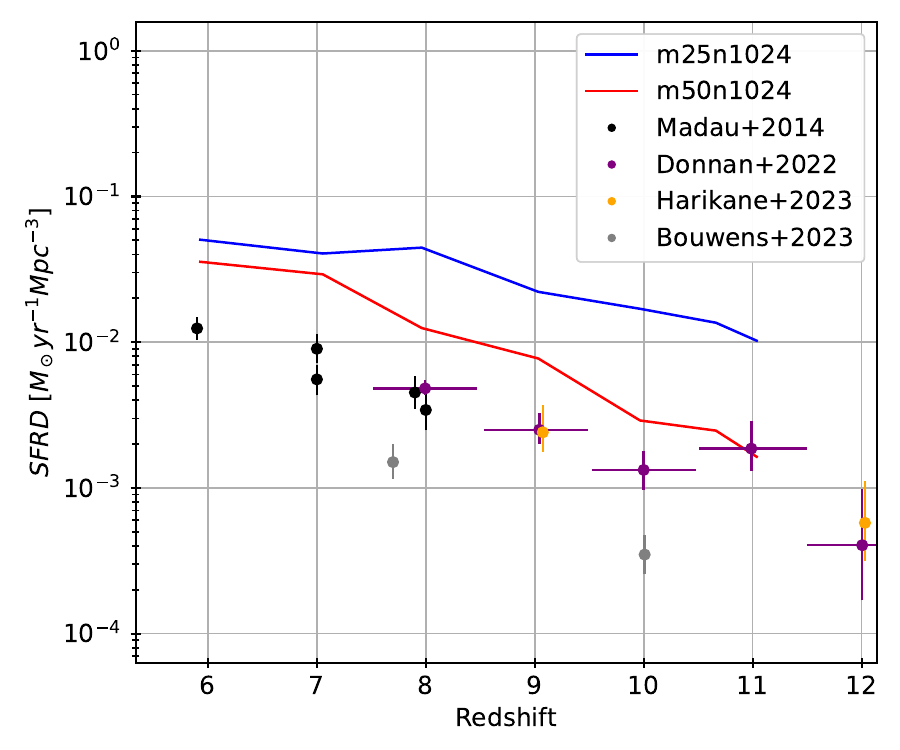}
        \caption{The SFRD's evolution from redshift $\sim 11$ to $\sim 6$. Red and blue lines denote our 50 Mpc/h and 25 Mpc/h runs respectively, whilst points show data from \citealt{bib:madau2014} (black), \citealt{bib:donnan2022} (purple) and \citealt{bib:harikane2023_sfrd} (orange). We see that whilst both of our simulations over-predict the SRFD consistently, the 50 Mpc/h run is a vastly better fit to the data than the 25 Mpc/h run.}
        \label{fig:madau}
    \end{figure}

    Figure \ref{fig:madau} shows the evolution of the star formation rate density (SFRD) with redshift. Our 50 Mpc/h and 25 Mpc/h runs are plotted as the red and blue lines respectively. The black, purple and orange points denote data presented in \citealt{bib:madau2014}, \citealt{bib:donnan2022} and \citealt{bib:harikane2023_sfrd} correspondingly. It is clear to see that in general, our simulations predict larger SFRDs than expected from the data across the entire redshift domain.

    There exists, however, a striking difference between our simulations, with the 50 Mpc/h low-resolution run displaying a significantly closer fit to the data at high redshift than its 25 Mpc/h counterpart. Whilst both simulated boxes are in agreement to $\sim 0.2$ dex at redshift 6, at redshift 11 they displaying a disparity of $\sim 0.9$ dex; the larger box in 1$\sigma$ agreement with the recently observed data point of \citealt{bib:donnan2022}. We see that the 25 Mpc/h run's SFRD-$z$ scaling has a far shallower gradient than that of the 50 Mpc/h run, resulting in the former's significant disagreement with observed sources at high redshift.

    Given the tension displayed here between both of our simulations, it is sensible to conclude the presence of some resolution dependency in the star formation mechanisms at high redshift. However, for future implementations of our model it is important to highlight the potential need for resolution-dependent tuning in the star formation models -- none of which was carried out by us in the creation of this paper. Furthermore, simulations running to lower redshifts ($< 6$) will require monitoring in this regard: our findings on our model's low-redshift behaviour will be presented in a future paper.

    \subsection{The Kennicutt-Schmidt Relation}
    \label{sec:results_fiducialSimba:kennicutt-schmidt}

    \begin{figure}
        \centering
        \includegraphics[width=1.\columnwidth]{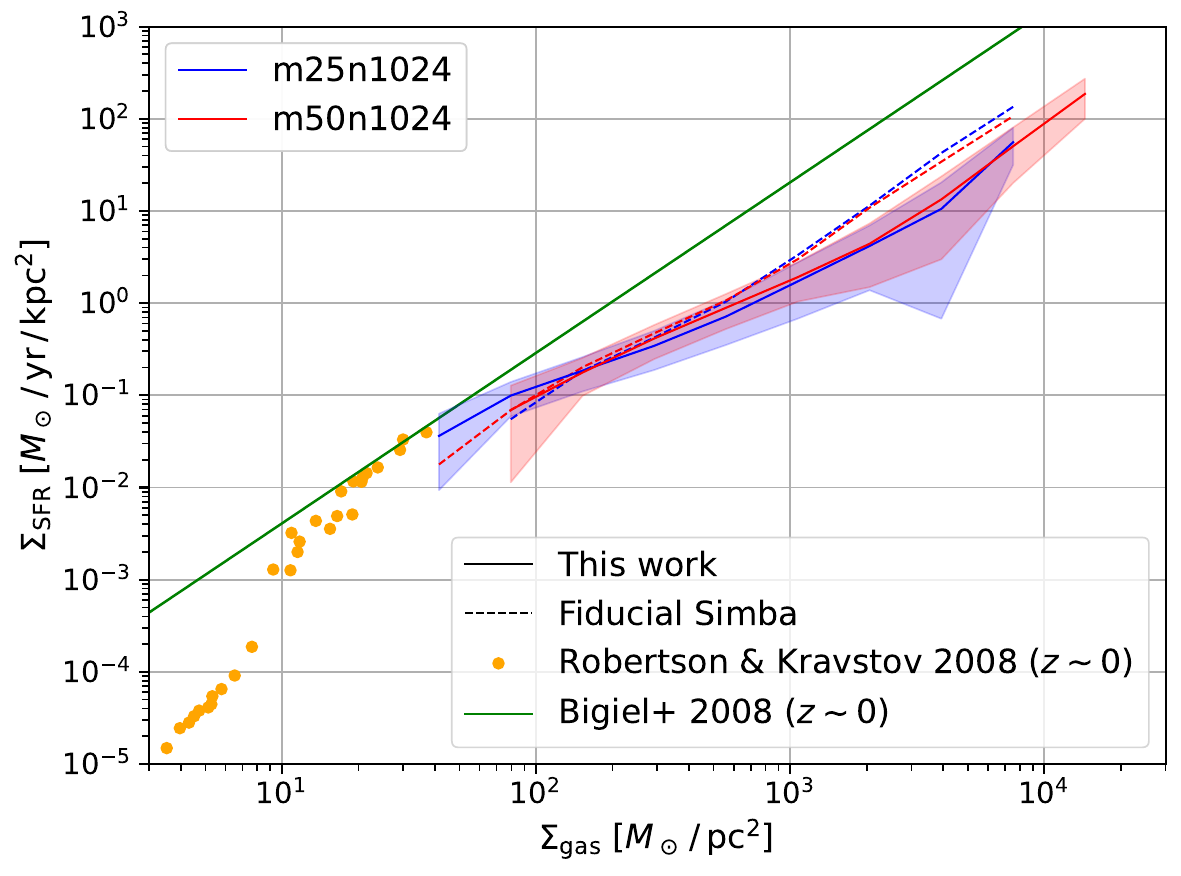}
        \caption{The median galactic SFR surface density vs gas surface density relation found using our model (solid lines) and fiducial \simba (dashed lines) at $z\sim6$. Blue and red colouring indicates results from the 25 Mpc/h and 50 Mpc/h boxes respectively. Furthermore, the orange data points show the $z\sim0$ datasets of \citealt{bib:robertson2008}, whilst the green line shows the power-law fit presented in \citealt{bib:bigiel2008}.}
        \label{fig:kennicutt-schmidt}
    \end{figure}

    In Figure \ref{fig:kennicutt-schmidt} we show the median Kennicutt-Schmidt (KS) relation (galactic SFR surface density vs gas surface density) for $z\sim6$ galaxies simulated using the models developed in this work (solid lines) and those of fiducial \simba (dashed lines) -- with blue/red denoting the 25/50 Mpc/h boxes. Shaded regions indicate the 1-$\sigma$ error within each $\Sigma_\mathrm{gas}$ bin. We plot the results of \citealt{bib:robertson2008} as orange datapoints, though analytical care must be taken as this is a $z\sim0$ sample. Furthermore, we plot the power-law fit presented in \citealt{bib:bigiel2008} as a green line; though this is a fit to local universe observations and so does not represent objects at the redshift of our simulation. Moreover, this fit does not contain objects of $\Sigma_\mathrm{gas} \gtrsim 10^{2.5}\,\,M_\odot/\mathrm{pc}^2$ in the original work, necessitating further caution when making comparison against the dense objects present in our simulation.

    Comparing the results of our model with that of fiducial \simba we see that both are in close agreement as to the magnitude and gradient of the KS relation. In the $\Sigma_\mathrm{gas} \lesssim 10^3\,\,M_\odot/\mathrm{pc}^2$ region our model is shown to reproduce the results of fiducial \simba to within 1-$\sigma$. Though these lines slowly diverge as the galactic mass increases they remain definitively within 2-$\sigma$. Furthermore, we do notice that the gradient of the KS relation found with \simbaeor appears to be increasing at the most gas-dense galaxies, resulting in a gradient comparable with that of the fiducial simulation. We observe no prominent difference in behaviour between the two simulated resolutions, with the results of each model only differing in their minimum/maximum $\Sigma_\mathrm{gas}$ values, whilst throughout their evolution the two boxes remain tightly coupled.

    The plotted data of \citealt{bib:robertson2008} contains results at lower gas densities than found in our simulations, with no region of overlap between the two. This is simply a result of the largely disparate redshifts between our simulations ($z\sim6$) and the data ($z\sim0$). However, at $\Sigma_\mathrm{gas} \sim 4\times10^1\,\,M_\odot/\mathrm{pc}^2$, we see that the galaxies simulated in our 25 Mpc/h box are in excellent agreement with the \citealt{bib:robertson2008} data. Moreover, extrapolating our results down to lower densities, we see that the high-density end of the dataset agrees to within $1-\sigma$ of \simbaeor's predictions. However, this extrapolated agreement must only be considered for $\Sigma_\mathrm{gas} \gtrsim 2\times10^1\,\,M_\odot/\mathrm{pc}^2$ as the data is shown to change gradient below this.

    The power-law fit provided by \citealt{bib:bigiel2008} shows good agreement with our simulations at $\Sigma_\mathrm{gas} \sim 10^{1.5}\,\,M_\odot/\mathrm{pc}^2$, with disparities between the two increasing until $\Sigma_\mathrm{gas} \sim 10^{2.5}\,\,M_\odot/\mathrm{pc}^2$, after which the fit predicts the SFR $\sim1$ dex greater than our simulations. It is important to highlight that the fit as originally presented does not increase past $\Sigma_\mathrm{gas} \sim 10^{2.5}\,\,M_\odot/\mathrm{pc}^2$ and therefore this order-of-magnitude discrepancy is inconclusive. Moreover, the fit is constructed from objects in the local universe, which will be at significantly lower redshifts than our simulations -- another reason to apply analytical caution.
    
    To make any conclusive statements on our model's performance with comparison to these datasets would be unwise with such a disparity in redshift (in addition to the density domain of the fit and simulations only coinciding for a small subset of sources), but the tight coupling between \simbaeor and \simba suggests that our model is performing at least as well as the fiducial model. Once we are able to run \simbaeor to lower redshifts, many observational datasets will become available for comparison.

    \subsection{The Mass-Metallicity Relation}
    \label{sec:results_fiducialSimba:massMetallicityRelation}

    \begin{figure*}
        \centering
        \includegraphics[width=1.6\columnwidth]{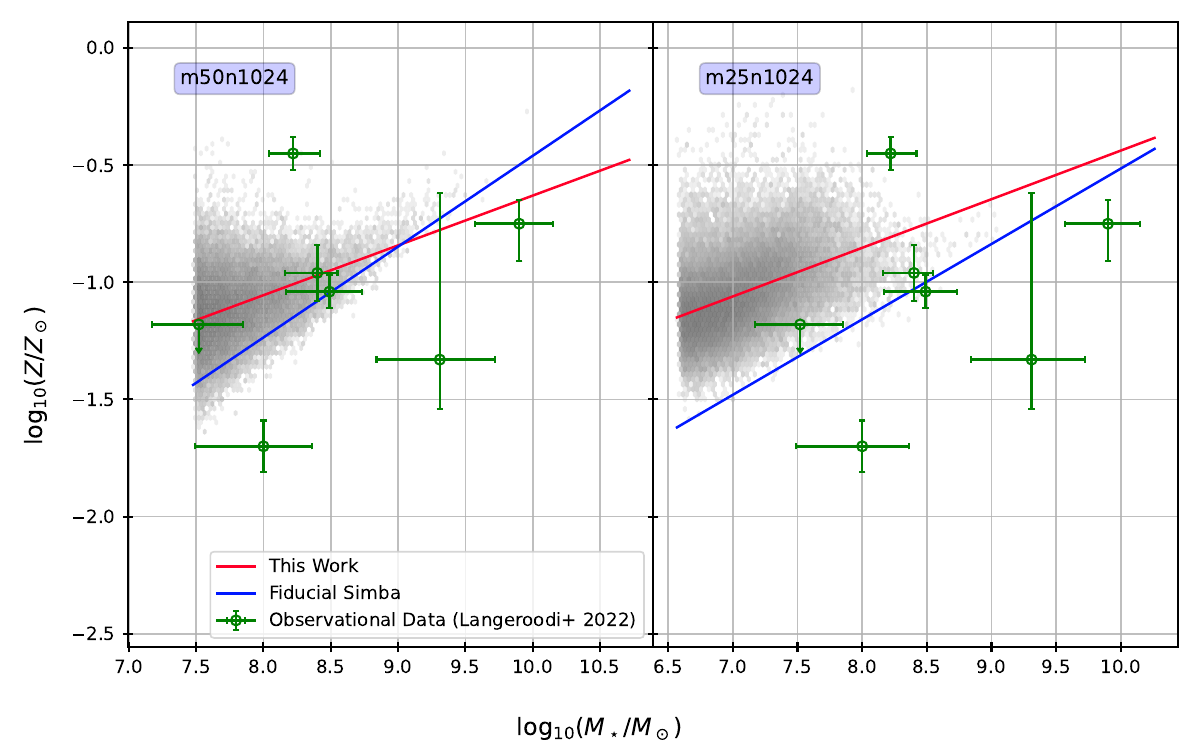}
        \caption{The mass-metallicity relation observed in both the 50 Mpc/h (left panel) and 25 Mpc/h (right panel) simulation boxes at $z\sim6$. Red and blue lines depict linear, least-square fits to the results produced by the models presented in this work and fiducial \simba respectively. Additionally, results from Table 3 in \citealt{bib:langeroodi2022} are plotted in green; these sources have spectroscopic redshifts in the range 7.2-8.5. Furthermore, the hexbin plot on each axis shows the distribution of galaxies found in our simulations, with darker shades corresponding to a higher count. Our models display more resolution dependence than the fiducial; the metallicity at fixed stellar mass increasing by $\sim 0.3$ dex in the high resolution box. Furthermore, the larger box shows considerably better agreement with the observational data plotted than its smaller counterpart.}
        \label{fig:massMetallicityRelation}
    \end{figure*}

     \begin{table}
        \centering
        \label{table:massMetallicityRelation}
        \caption{Results of the linear (in log-space) least-squares fits for the mass-metallicity relation presented in Figure \ref{fig:massMetallicityRelation}.}
        \begin{tabular}{llll}
            \hline
            Box-size (Mpc/h) & Model & Gradient & Intercept\\
            \hline
            50 & This Work & $(2.13 \pm 0.022)\times10^{-1}$ & $-2.8 \pm 0.017$ \\
            50 & Fiducial & $(3.9 \pm 0.030)\times10^{-1}$ & $-4.3 \pm 0.024$ \\
            25 & This Work & $(2.1 \pm 0.017)\times10^{-1}$ & $-2.5 \pm 0.012$ \\
            25 & Fiducial & $(3.2 \pm 0.047)\times10^{-1}$ & $-3.7 \pm 0.034$ \\
        \end{tabular}
    \end{table}

    In Figure \ref{fig:massMetallicityRelation} we plot show the mass-metallicity relation in the 50 Mpc/h and 25 Mpc/h boxes (left and right panels respectively) for both the simulations presented in this work and fiducial simba: these are shown in red and blue respectively. The plotted lines show a least-squares, linear fit (in log-space) to the underlying data, the results of which are detailed in Table \ref{table:massMetallicityRelation}. We also add a subset of the observational data presented in Table 3 of \citealt{bib:langeroodi2022} as green points on the plot. We choose these points such that the observed sources have a spectroscopic redshift in the range $7.2 \lesssim z \lesssim 8.5$. Lastly, we show the distribution of galaxies identified in our simulations on the hexbin in the background. This is shaded by the number of counts in each bin, with darker shades indicating a larger number of sources.

    Comparing the two panels we immediately notice the shift toward higher metallicities introduced in our simulations upon increasing the resolution. In contrast to our 50 Mpc/h fit, which is seen to lie within within 1$\sigma$ of many observational points, our 25 Mpc/h fit seems to consistently overestimate the metallicity at given stellar mass. From Table \ref{table:massMetallicityRelation} we see that the 25 Mpc/h run has an intercept $\sim 0.3$ dex higher than its low-resolution counterpart. This, coupled with both runs having very similar gradients which are within 0.03 dex of one another, leads to the metallicity being systematically predicted $\sim 0.3$ dex larger in the 25 Mpc/h box than the 50 Mpc/h box.

    In contrast, comparing the fits of the fiducial runs we see different behaviour; namely their gradients being dissimilar, with the fit for the 50 Mpc/h box possessing a gradient $\sim 20\%$ larger than its 25 Mpc/h equivalent. In the 50 Mpc/h box, the difference in gradient between the fit of our model and fiducial \simba results in their coincidence at $\logten(M_\star/M_\odot) \approx 9$. We see that the fiducial run fits the observational data in the 50 Mpc/h box less-well than our run. However, the opposite appears true for the 25 Mpc/h box, where the fiducial run's fit does not yet intercept ours, staying below it and passing through the observational sources which our models overestimated.

    The discrepancy between boxes discussed does seem to be manifestation of resolution dependence within the simulations, it instead may arise from the tuning (or lack thereof) within our models or the star formation model. Sub-grid models such as these require calibration of their free parameters to produce consistent results across different resolutions, which is usually achieved through comparison with known results, and is commonplace when using such models. Given that no such tuning was undertaken for this paper, we do believe that Figure \ref{fig:massMetallicityRelation} highlights a lessened resolution-dependency in our models than those of fiducial \simba. Although our runs perform worse (compared to the observational data provided) than the fiducial in the 25 Mpc/h box, what we do see is a consistency between the functional form of our fits which is not present in the fiducial. Whilst the fractional errors on the gradient and intercept between the larger and smaller boxes in our runs are 2.4\% and 8.8\% respectively. The analogous results for the fiducial runs are a fractional error on the gradient of 17\% and intercept of 14\%. The smaller fractional errors observed above for our runs are a clear indication that integration of our models has weakened the resolution dependency of the simulation and therefore improved the stability of physical results.

\section{Model Predictions: Comparison to KMT}
\label{sec:results_kmtComparison}

With confidence that our simulations are a reasonably accurate match to observed astrophysical relations, as shown in Section \ref{sec:results_fiducialSimba}, here we discuss specifically the modelling differences which arise from our replacement of the KMT model with our new two-phase ISM model.

To achieve a direct comparison we have taken a sample of 100 galaxies at four stellar masses from $10^7 - 10^{10} \,\, M_\odot$. For each SPH particle within a selected galaxy, we have computed the H$_2$ fraction as estimated by the KMT model \citep{bib:KMT2009}. This was done by constructing a ray centred on each particle, extending $\pm1$ smoothing length along the $x$-axis, and integrating the dust mass along it to calculate the optical depth. This is slightly different to how it is calculated at \simba run-time, which uses the Sobolev approximation, but still serves as a reasonable calculation given the lack of differential fields in the snapshot. Using the dust optical depth we then calculate the molecular hydrogen fraction as prescribed by KMT \citep{bib:KMT2009, bib:KG2011}. Post-processing in this way allows is to compare between the explicit chemical modelling and KMT approximation for identical particles -- something which cannot be achieved through comparison of simulated runs using both models (such as that presented in Section \ref{sec:results_fiducialSimba}).

    \subsection{The Distribution of Molecular Hydrogen}
    \label{sec:results_kmtComparison:histogram}

    \begin{figure}
        \centering
        \includegraphics[width=\columnwidth]{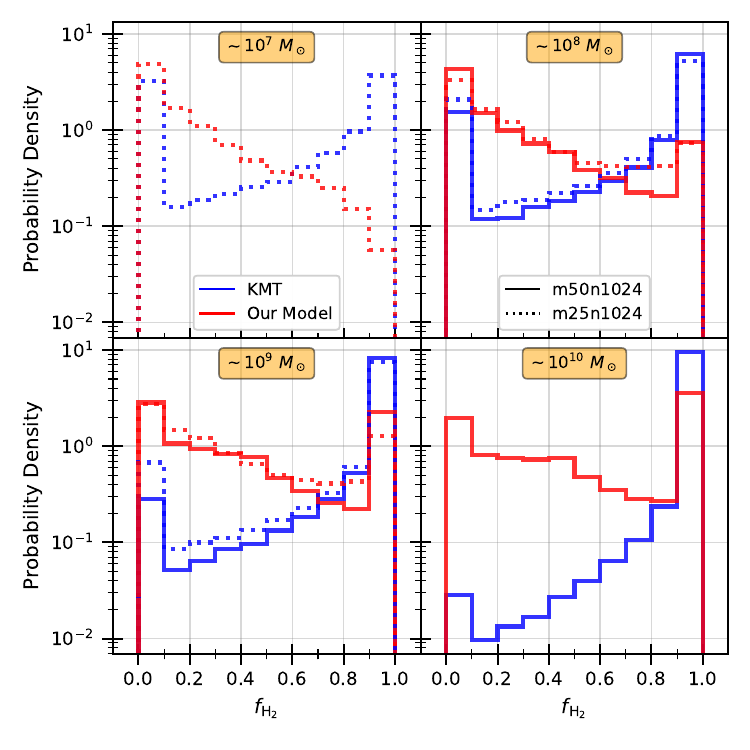}
        \caption{A histogram showing the probability density for a galactic SPH particle to contain a given fraction of H$_2$. The red lines show the molecular hydrogen fraction calculated using our new model, the solid and dotted lines representing data from the 50 and 25 Mpc/h boxes respectively. The blue lines show a post-processed calculation of the H$_2$ fraction as estimated by the KMT model for each particle in our dataset. We selected the 100 galaxies with stellar masses closest to $10^7,10^8,10^9$ and $10^{10} \,\, M_\odot$ to create four datasets sampling the stellar masses present in both boxes; the stellar mass of the sample is shown as an annotation on its axis. It is seen that the KMT model consistently predicts large amounts of fully molecular gas throughout the stellar mass range, whereas our model shows a procedural rise in the population of fully molecular particles with increasing stellar mass. Furthermore, the KMT model displays significantly less partially-molecular gas particles than our model, an effect which is exacerbated at large stellar masses, where we see that our model predicts a much flatter distribution.}
        \label{fig:fH2_kmtComparison}
    \end{figure}

    Figure \ref{fig:fH2_kmtComparison} shows histograms of the probability density of a galactic SPH particle possessing a H$_2$ fraction within one of ten equally-spaced bins. The red and blue lines show the molecular hydrogen fractions as calculated by our model during simulation run-time, and as calculated from the KMT model through post-processing, respectively. The solid lines represent results for the 50 Mpc/h box, whilst the dotted lines represent the 25 Mpc/h box. Each subplot contains the 100 closest galaxies to the stellar mass annotated on the axis, allowing us to cover a wide and representative range of galaxies in our analysis.

    Focusing first on the sample of $\sim 10^7 \,\, M_\odot$ galaxies, we immediately see that our model and KMT have major disagreements when $f_{\mathrm{H}_2} > 10 \%$. Whilst the KMT model predicts roughly the same amount of fully molecular as fully atomic gas (we use the terms fully molecular and fully atomic in this section to refer to the highest and lowest bins respectively, for convenience) our simulation predicts almost no fully molecular, and an abundance of fully atomic, gas. For particles with $f_{\mathrm{H}_2} > 10 \%$ we see a positive correlation with the probability density in the KMT case, and a negative correlation for our model.

    As we increase the stellar mass of the galaxies in our sample, we see shifts in the behaviour mentioned for the $\sim 10^7\,\,M_\odot$ case. In the $\sim 10^8\,\,M_\odot$ sample, the KMT model is predicting significantly more particles in the fully molecular regime than the fully atomic, meaning the probability density for a particle to contain $f_{\mathrm{H}_2} < 90 \%$ falls significantly. Our models show the opposite behaviour, the population of fully molecular particles increasing dramatically, flattening the slope of the intermediate values as a result.

    A continuation of these behaviours is seen in the $\sim 10^9\,\,M_\odot$ sample, the KMT model again shifting particles towards the fully molecular and away from the fully atomic. We see here that a particle pulled from the KMT model's distribution is over $10\times$ more likely to possess $90\% < f_{\mathrm{H}_2} \leq 100\%$ than the next most probable bin. On the other hand, our model's distribution is flattening, and though the population of fully molecular gas continues to increase, we do not see the exceptionally low relative probabilities exhibited in the KMT case.

    Finally, the histogram containing the $\sim 10^{10} \,\, M_\odot$ stellar mass sample shows an extension of the effects present before. Here, the KMT model shows an exceptionally steep gradient in the probability density between its two highest bins, increasing by almost two orders of magnitude. The lower bins, including the once significant fully atomic bin are roughly $1000\times$ less likely than the fully molecular case. Conversely, the histogram for our model shows the fully molecular bin as containing only $2\times$ the probability density of its atomic counterpart. This is a significant difference between the models, KMT predicting that the vast majority of particles in massive galaxies are fully molecular. The other important distinction between the models here are their gradients. As mentioned, KMT shows a very large spike at the highest bin, and has a rapidly increasing gradient prior. Our model, however, displays a rather flat distribution, the probability density varying by roughly 0.5 dex in the range $10\% < f_{\mathrm{H}_2} \leq 90\%$. Whilst these intermediate bins are still less probable than the extremes, we would reasonably expect to see a large number of particles containing these fractions in our simulations; the same cannot be said for those in the KMT model.

    The last observation to discuss in Figure \ref{fig:fH2_kmtComparison} is the agreement between the simulated box sizes. Looking at the intermediate stellar mass bins (the highest and lowest contain particles from only one of the boxes due to their resolution limits) we see that in the case of both the KMT and our model, the distributions for both box sizes are in good agreement. Whilst there are discrepancies in the exact probabilities themselves, the general trends discussed above are clearly followed at both resolutions.

    \subsection{The H$_2$ Fraction's Metallicity and Dust Scaling}
    \label{sec:results_kmtComparison:H2_metDust}

    \begin{figure}
        \centering
        \includegraphics[width=1\columnwidth]{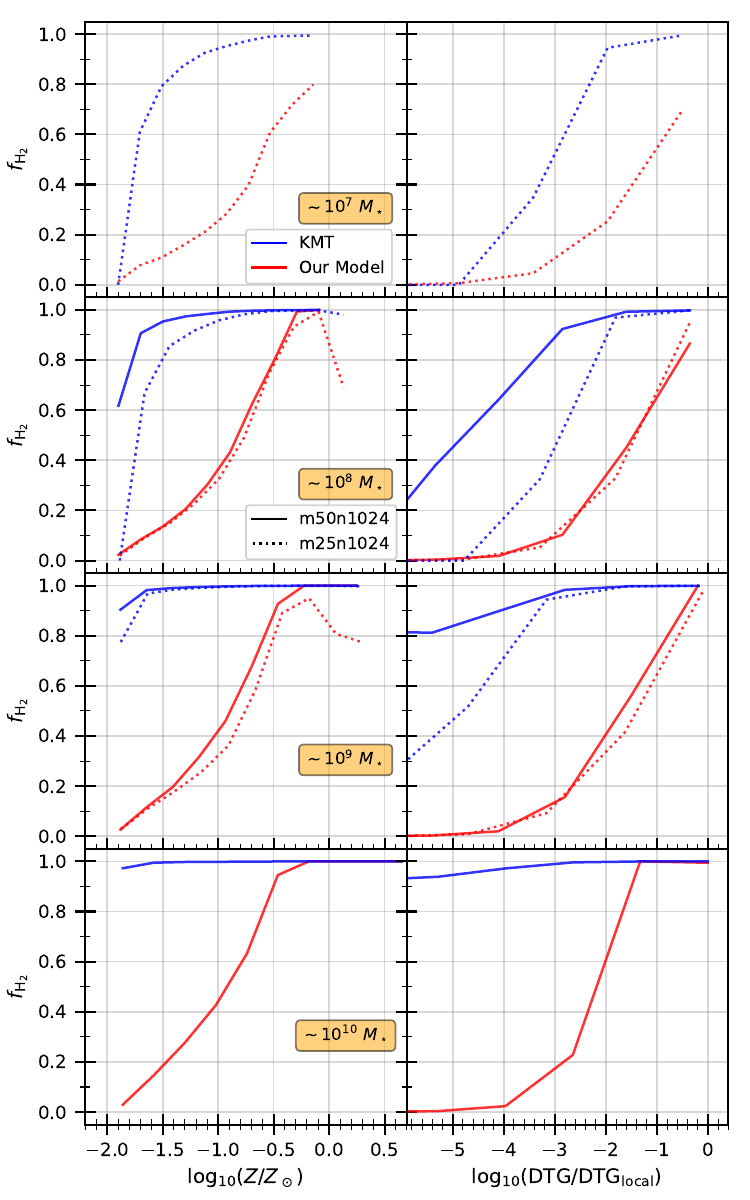}
        \caption{The dependence of metallicity and dust-to-gas ratio on the H$_2$ fraction in galaxies of varying stellar mass. In each row of the figure we show results for a sample of the 100 galaxies closest to the stellar mass indicated in the annotation in the left panel: these samples are identical to those presented in Figure \ref{fig:fH2_kmtComparison}. The left and right columns of the plot show how $f_{\mathrm{H}_2}$ varies with the metallicity and dust-to-gas ratio (normalised to the Milky Way value of \citealt{bib:pollack1994}) respectively. In each panel the x-axis is split into 10 equally-spaced bins, the median molecular hydrogen fraction within each calculated. Red and blue lines show results from our model and the KMT model respectively, with solid and dashed lines distinguishing between the simulated box-sizes as denoted on the figure.}
        \label{fig:H2_metDtg}
    \end{figure}

    Continuing our investigation into the modelling of molecular hydrogen, Figure \ref{fig:H2_metDtg} shows how the median H$_2$ fraction scales with key physical quantities across ten equally-spaced bins. In the left column we show how metallicity affects $f_{\mathrm{H}_2}$, and in the right column, its scaling with the dust-to-gas ratio, which we normalise to its Milky-Way value of $9.387\times10^{-3}$ \citep{bib:pollack1994}. Each row of the figure represents a different sample of 100 galaxies around the stellar mass indicated by the annotation in the left panel. These samples are identical to those used in Figure \ref{fig:fH2_kmtComparison}. Blue and red lines show results from the KMT model and our model respectively, the solid and dotted lines distinguishing between the simulated box-sizes.

    The top row of the figure plots the scaling relations for galaxies with stellar masses $\sim 10^7\,\,M_\odot$, on which we see an immediate disparity between the models. Whilst the molecular hydrogen content of the gas increases steadily with metallicity in our model, the KMT model shows an extremely rapid scaling, reciprocal in nature. As the metallicity increases from $Z = 10^{-2}\,\,Z_\odot$ to $Z = 10^{-3/2}\,\,Z_\odot$ in the KMT model, the median molecular hydrogen fraction increases from $0\%$ to $80\%$ whilst our model reaches only $10\%$ over this interval. At this stellar mass we see that the KMT model reaches the fully molecular regime at $Z \approx 10^{-1/2}$, at which point our model predicts that only $60\%$ of the hydrogen is molecular. Similar behaviour is seen when considering how the H$_2$ fraction scales with the dust content of the gas. The KMT model again predicts a stronger dependence on the dust-to-gas ratio when it comes to forming molecular hydrogen than our model. This is seen clearly in the gradients of the curves, and the maximal H$_2$ fraction reached in each case.

    The $10^8\,\,M_\odot$ sample shows much the same behaviour as the $10^7\,\,M_\odot$ sample; the KMT model exhibiting an extremely rapid increase in molecular hydrogen content with metallicity, plateauing into the fully molecular regime before our model predicts a H$_2$ fraction of even $50\%$. A difference we do see to the previous sample however, is that our model does reach the fully molecular regime at sufficiently large metallicities. Though the gradient of the curve is much the same in both samples, the increased galactic mass has allowed for higher metal and dust content. In the KMT model, the 25 Mpc/h box scales at much the same rate in both samples also. Furthermore, we see that whilst the difference between the box-sizes in our model are generally minimal (except at the highest metallicities where resolution effects are seen), the KMT model has major disparities in both the metallicity and DTG scaling relations. For example, the 50 Mpc/h box predicts $f_{\mathrm{H}_2} \approx 60\%$ at $Z \approx 10^{-1.9}\,\,Z_\odot$, whereas the 25 Mpc/h box predicts that the hydrogen is fully atomic. A similar, though less exaggerated discrepancy can be see in the DTG scaling, where the larger box forms molecular hydrogen much more readily than its smaller counterpart.

    As we move to the $10^9\,\,M_\odot$ sample we see the KMT model is now predicting much more molecular hydrogen than it was previously. Where the 25 Mpc/h box's scaling relations didn't change too much between the previous two samples, here we see a dramatic shift towards the fully molecular regime. In the left panel we see the KMT model predicts $f_{\mathrm{H}_2} > 70\%$ at even the lowest metallicities, whereas our models are predicting $f_{\mathrm{H}_2} < 5\%$. Furthermore, the right panel also shows a distinct increase in the KMT model's gradients compared to the last sample; though this is not as pronounced as it is in the left panel. On the contrary, we see that our model, in both its metallicity and DTG scaling, displays congruent behaviour with the previous stellar mass bin, whilst showing very little resolution dependency.

    The final sample shows almost no scaling with either the metallicity of dust-to-gas ratio for the KMT model, its hydrogen remaining $>90\%$ molecular in the least dusty and metallic conditions. Looking back to Figure \ref{fig:fH2_kmtComparison}, we see this behaviour manifest in the radical increase in probability between the highest two $f_{\mathrm{H}_2}$ bins. On the other hand, the metallicity scaling for our model is seen to remain roughly consistent with its previous behaviours at lower galactic stellar masses; whilst the gradient of the DTG scaling is seen to increase in dustier particles. Examining again the other samples, we do see that the $f_{\mathrm{H}_2}$ vs DTG curve for our model steepens gradually with the stellar mass of the sample. However, the same cannot be said for the metallicity relation, which remains very stable, ignoring resolution effects. Relative to the strong dependence on the stellar mass that the KMT model exhibits here, our model is largely independent of the size of the host galaxy.

    \begin{figure}
        \centering
        \includegraphics[width=\linewidth]{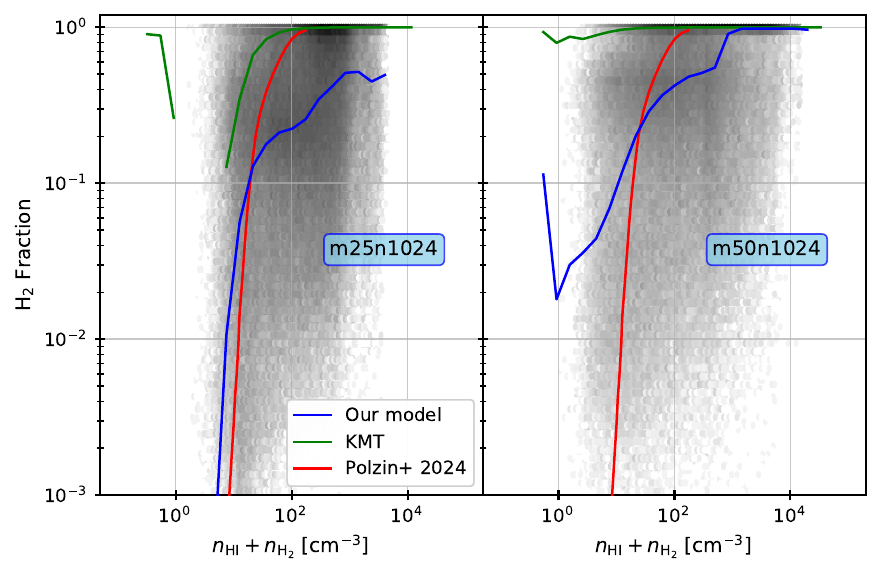}
        \caption{The median molecular hydrogen fraction of galactic particles vs the neutral hydrogen density. We include all galactic particles that exist within a sample of 300 simulated galaxies, chosen identically as in Figures \ref{fig:fH2_kmtComparison} and \ref{fig:H2_metDtg}. The blue, red and green lines show the results for our model, KMT and \citet{bib:polzin2024} respectively. Beneath the median lines we show a hexbin plot for the count of particles within each bin, darker cells correspond to a larger count.}
        \label{fig:fH2_hydrogen}
    \end{figure}

    Complementing this analysis, we also show the molecular hydrogen fraction as a function of the $\mathrm{HI} + \mathrm{H}_2$ number density in Figure \ref{fig:fH2_hydrogen}. The blue, green and red lines show the median H$_2$ fractions from our \simbaeor model, post-processed KMT estimations and the results of \citet{bib:polzin2024} respectively. For each box-size we show all particles within a 300 galaxy sample, 100 galaxies at each stellar mass as shown in Figures \ref{fig:fH2_kmtComparison} and \ref{fig:H2_metDtg} -- although we do not distinguish between the bins in this case. Behind the median lines we display a monochromatic hexbin distribution showing the particle count within each bin; darker cells containing a higher number of particles, though the exact numbers are not important for our analysis. However, it is important to note that the simulations of \citet{bib:polzin2024} are explicitly for the ISM of a galactic disk, which is not only significantly higher resolution, but has limited context when compared with the diverse sample of galaxies present in our cosmological simulation.

    Our model shows reasonably good agreement with the results of \citet{bib:polzin2024}, particularly in the small box at densities associated with the sharp transition from atomic to molecular gas ($n \sim 10 \,\, \mathrm{cm}^{-3}$). In both our boxes, the slope of $f_{\mathrm{H}_2}$ vs. $n_\mathrm{HI} + n_{\mathrm{H}_2}$ becomes significantly shallower above $f_{\mathrm{H}_2} \sim 0.2$, whereas in \citet{bib:polzin2024}, $f_{\mathrm{H}_2}$ reaches unity within a dex of the transition density. This could highlight differences in the physical conditions at high density experienced in our cosmological simulation compared with the single galaxy of \citet{bib:polzin2024}.
    
    Comparing our results to the KMT model, we see that KMT estimates larger H$_2$ fractions across the density domain -- this is expected given the results of our previous plots (see Figures \ref{fig:fH2_kmtComparison} and \ref{fig:H2_metDtg}). Finally, we also see, particularly in the large box, a rise in the median molecular hydrogen fraction at very low densities. This is due to a small fraction of dusty particles that exist within newly star-forming regions. We discuss this small population of particles fully in Section \ref{sec:results_dustAndISM:ISRF} and reserve further discussion until then.

\section{Model Predictions: Dust and ISM Properties}
\label{sec:results_dustAndISM}

In this section we investigate the dust and ISM properties calculated using our new model. To achieve this we added four new fields to the snapshot outputs written by \simba: the dust temperature, the local interstellar radiation field strength incident on the particle, and both the density and internal energy of the cool ISM component added in our two-phase model. Here we will explore the relationship between these and verify that our modelling is realistic and produces sensible, physically consistent results.

    \subsection{The Dust-to-Gas and Dust-to-Metal Ratios}
    \label{sec:results_dustAndISM:dust-to-gas/metal}

    \begin{figure*}
        \centering
        \includegraphics[width=1.4\columnwidth]{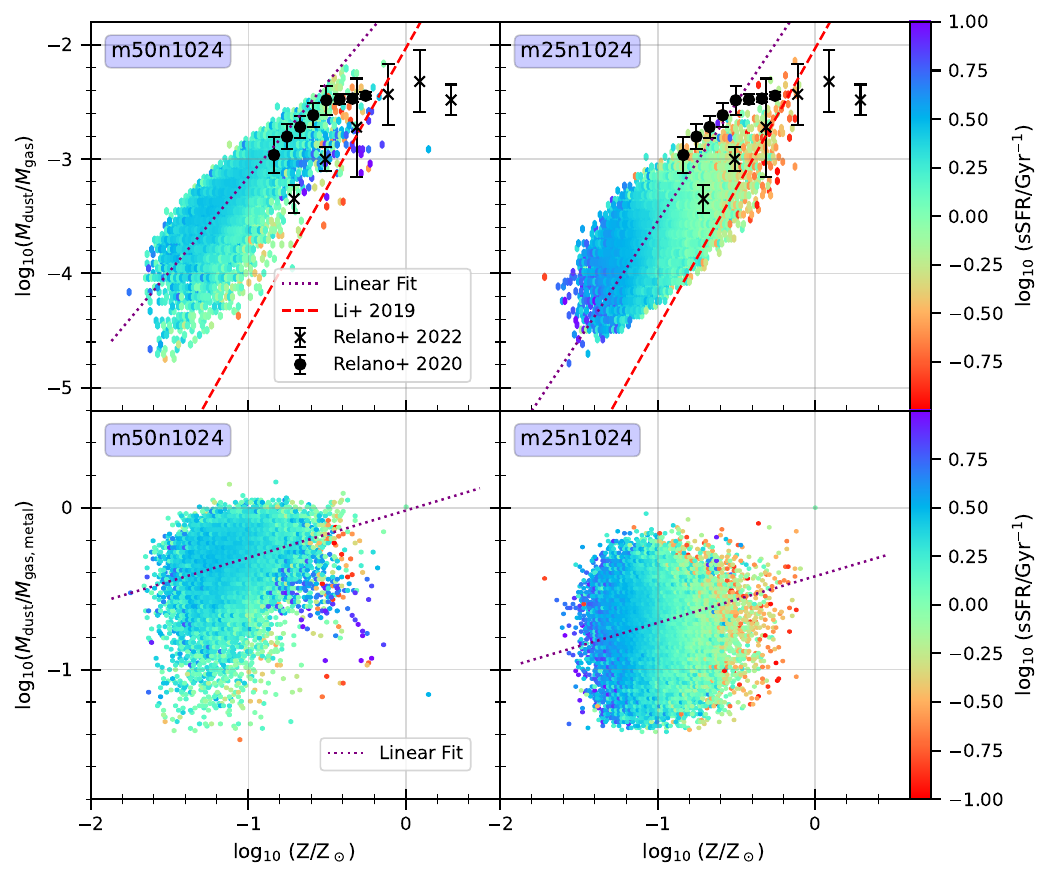}
        \caption{The dust-to-gas ratio (top row) and dust-to-metal ratio (bottom row) as a function of metallicity for galaxies identified in both our 50 (left column) and 25 (right column) Mpc/h boxes. The colour bar shows the mean specific star formation rate in each bin. The dotted purple lines are linear (in log-space) least-squares fits through the data. The red dashed line present in the top row shows the linear fit from \citealt{bib:Li2019}. Additionally, we show data collected from infrared observations; crosses depicting the results presented in \citealt{bib:relano2022}, and circles depicting the results presented in \citealt{bib:relano2020} for the M101 galaxy. It is clear to see that the dust-to-gas ratio from our model grows less-steeply than that of its predecessor, and that our simulated galaxies contain dust populations comparable to those observed.} For the dust-to-metal ratios, we see that our fits highlight a relatively flat, though positively correlated dependence on metallicity.
        \label{fig:dtgm_z}
    \end{figure*}

    \begin{figure}
        \centering
        \includegraphics[width=\columnwidth]{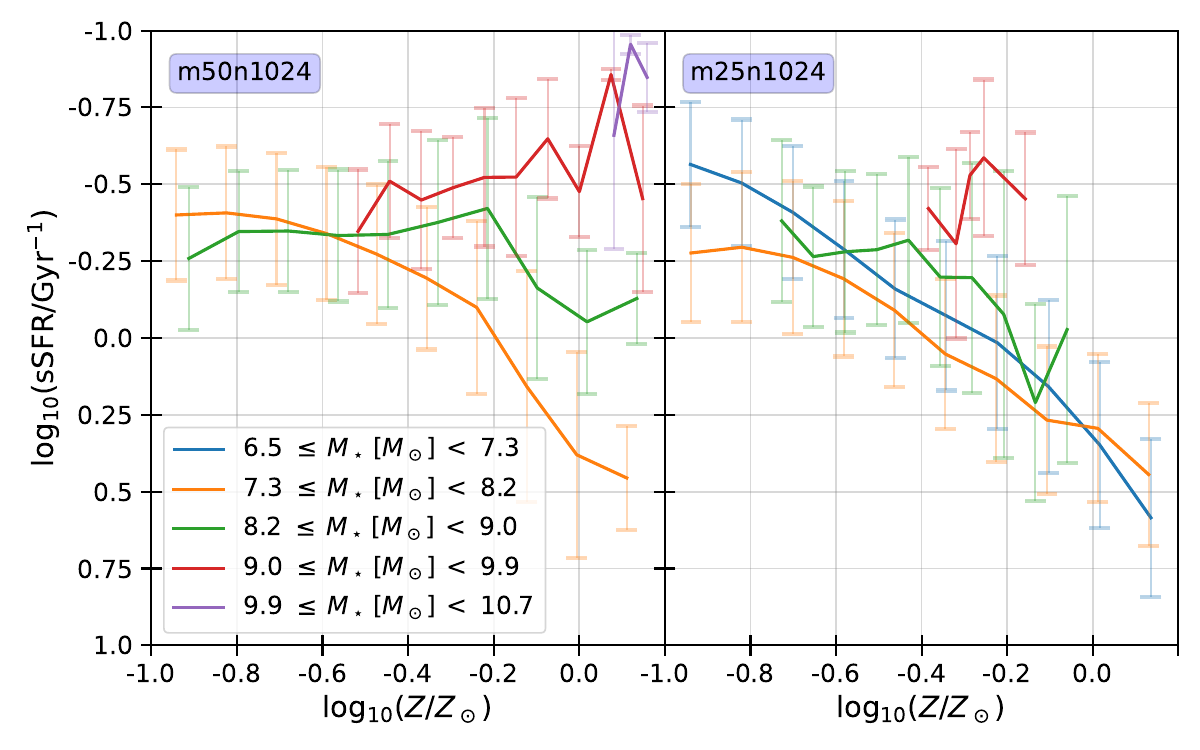}
        \caption{The median sSFR as a function of metallicity for all galaxies in each simulation box. The coloured lines (described for each axis individually in the legend) show the dependency of galactic stellar mass on the slope of the sSFR-Z relation. Error bars show the 1-$\sigma$ deviation within each metallicity bin, each of which are required to contain three galaxies at minimum to appear on the plot. This constraint is only relevant for the most massive galaxies in each box as they are sparse in number. It can be seen that the low-mass galaxies exhibit decreasing sSFR with increasing metallicity, whereas the largest galaxies (which are found primarily in the 50 Mpc/h box) show a positive sSFR-Z correlation. This behaviour can explain the outlying points of high metallicity and sSFR in Figure \ref{fig:dtgm_z}.}
        \label{fig:ssfrMetallicity}
    \end{figure}

    \begin{table*}
        \centering
        \label{table:dgr_fits}
        \caption{The mathematical expression of each best fit line for the DGR as a function of metallicity -- as seen in Figure \ref{fig:dtgm_z}.}
        \begin{tabular}{lcc}
            \hline
            Simulation & Redshift & Fit\\
            \hline
            \citet{bib:Li2019} & 0 & $\log_{10}\left(\mathrm{DGR}\right) = \left(2.445 \pm 0.006\right)\log_{10}\left(Z/Z_\odot\right) - \left(2.029 \pm 0.003\right)$\\
            This work: 50 Mpc/h box & 6 & $\log_{10}\left(\mathrm{DGR}\right) = \left(1.405 \pm 0.005\right)\log_{10}\left(Z/Z_\odot\right) - \left(1.779 \pm 0.006\right)$\\
            This work: 25 Mpc/h box & 6& $\log_{10}\left(\mathrm{DGR}\right) = \left(1.809 \pm 0.006\right)\log_{10}\left(Z/Z_\odot\right) - \left(1.752 \pm 0.007\right)$\\

        \end{tabular}
    \end{table*}

    In Figure \ref{fig:dtgm_z}, we find that the dust-to-gas ratio (DGR) in our model rises less steeply with metallicity than the original fit provided in \citealt{bib:Li2019}; both converging toward agreement at around one solar metallicity. At low metallicity, our model produces a significantly larger DGR than the fit of Li et al., showing disparity of $\sim1.3$ dex  at $Z = 0.1\,\, Z_\odot$ for the 50 Mpc/h box. The 25 Mpc/h box shows a result similar to, although slightly lower than, this with a $\sim0.9$ dex disparity at identical metallicity. This discord may be largely be explained by the fact that both the original fit and our fit were computed at disparate redshifts of  $z=0$ and $z=6$ respectively: Table \ref{table:dgr_fits} details the explicit functional form of each fit.
    
    As the metallicity approaches its solar value, $\logten(Z/Z_\odot) = 0$, both fits converge to a difference of only $\sim 0.3$ dex. Furthermore, the dust populations seen at high metallicities show good agreement with the $z\sim0$ observations presented in \citealt{bib:relano2020} and \citealt{bib:relano2022}; the former's data closer to our fit, and the latter's closer to \citealt{bib:Li2019}'s fit. As the redshift between our simulation and these observations is large it is unwise to make any conclusions further than the fact that our simulation produces galaxies with dust populations akin to those seen in reality.
    
    Additionally, we observe that for both boxes, the (average) sSFR falls with increasing metallicity. The 50 Mpc/h box however shows a peak in the sSFR at the high-metallicity and displays a generally weaker trend than the 25 Mpc/h box in this regard.

    The relationship observed in Figure \ref{fig:dtgm_z}, which shows a clear positive correlation between DGR and metallicity, is expected from physical considerations alone. We know that dust grows via accretion of metals (Section \ref{sec:methods:dustGrowthAndDestruction:metalAccretion}) and will therefore have higher abundance within gas of greater metal content. Inversely, gas with little metal will be unable to sustain high grain-accretion rates, suppressing the DGR ratio. However, there may exist physical environments in which the DGR is lower than expected for a given gaseous metal fraction. In such cases, as can be seen in Figure \ref{fig:dtgm_z} as the points lying well below the fit, although the metal content is sufficient to sustain a larger abundance of dust, there also exists a non-negligible contribution to the destruction of grains. For example, in highly star-forming regions it may be the case that dust destruction due to thermal sputtering (which becomes more efficient in dense environments) or shocks are more/equally as potent as the rate at which the dust is being produced/grown. This scenario would result in a DGR lower than expected from the metallicity of the gas alone.
    
    Additionally, from Figure \ref{fig:dtgm_z} we conclude that the sSFR decreases with increasing metallicity; as can be seen most clearly in the higher-resolution 25 Mpc/h simulation. Although this result can seem somewhat perplexing upon first inspection, it is a well-observed (see \citealt{bib:andrews2013} for example), though non-universal relationship. In \citealt{bib:yates2012}, the slope of the metallicity-SFR relation is shown to vary with galaxy mass, becoming positive/negative in high/low-mass galaxies: which we explore further in Figure \ref{fig:ssfrMetallicity}. Here we observe an obvious negative sSFR-Z correlation within low-mass galaxies, which turns-over at galactic stellar masses $M_\star \gtrsim 10^9 M_\odot$ to exhibit a positive correlation; as described in the aforementioned \citealt{bib:yates2012}. Through comparison of both figures, we can deduce that the points of largest sSFR (which peak at $\sim 10 \,\, \mathrm{Gyr}^{-1}$) in the 50 Mpc/h box are contributions from the most massive galaxies. It is clear then, as to why the negative sSFR-Z correlation depicted on the colour-bar of Figure \ref{fig:dtgm_z} is more pronounced within the 25 than the 50 Mpc/h box -- there is a lack of galaxies massive enough to have entered the positively-correlated regime in the smaller box.
    
    A complex and multi-faceted analysis is required for a complete physical understanding of an individual galaxy's placement on the DGR-Z-sSFR plane. Unfortunately, such physical discussion is beyond the scope of this paper and so we recommend reading \citealt{bib:yates2012} for a rigorous discourse on the subject; subsequently deferring any further analysis of our own to a future work.

    Complementing the above discussion on the dust-to-gas ratio, Figure \ref{fig:dtgm_z} also presents the dust-to-metal (DTM) ratio as a function of metallicity in the bottom row of panels: the dotted line showing a least-squares fit of the data to a linear function (in log-space). The fit for both boxes display a shallow, positive correlation with a gradient of $0.292 \pm 0.005$ and $0.288 \pm 0.004$ for the 50 and 25 Mpc/h boxes respectively. Although the DTM ratio scales at almost the same rate in both boxes, it is clear to see that there is a large offset between them at a given metallicity. For example, the larger box estimates an equal amount of dust and gas metals, whereas the smaller box's fit predicts $M_\mathrm{dust} \sim 0.4 M_\mathrm{gas,metal}$, at solar metallicity.
    
    The observed positive correlation between the DTM ratio and metallicity is expected from both the literature \citep{bib:deVis2019, bib:priestley2021, bib:Li2019} and the modelling done in our simulation. Equations \ref{eq:dustGrowth} and \ref{eq:dustGrowthTimescale} show that the rate of dust growth via accretion is directly proportional to the metallicity of the gas. Therefore, in high-$Z$ environments we observe considerable migration of metal from the gas to dust phase, increasing the DTM ratio.

   A result of particular note is that the DTM ratio exceeds unity in the 50 Mpc/h box, showing that a small subset of galaxies in our sample possess more metallic mass in dust than they do in gas: this is not observed in the smaller box. Furthermore, we see different shapes in the distributions between box sizes, the smaller box possessing a population of high-metallicity low-DTM galaxies which are largely absent in the larger box. The existence of this population in the smaller box further validates our discourse in Section \ref{sec:results_kmtComparison:H2_metDust}, where we argue that the abundance of dust does not necessarily increase with metallicity as its production is dependent upon stellar feedback mechanisms which are somewhat sensitive to the resolution of the simulation.

    \subsection{The Dust Temperature}
    \label{sec:results_dustAndISM:ISRF}

    \begin{figure*}
        \centering
        \includegraphics[width=2\columnwidth]{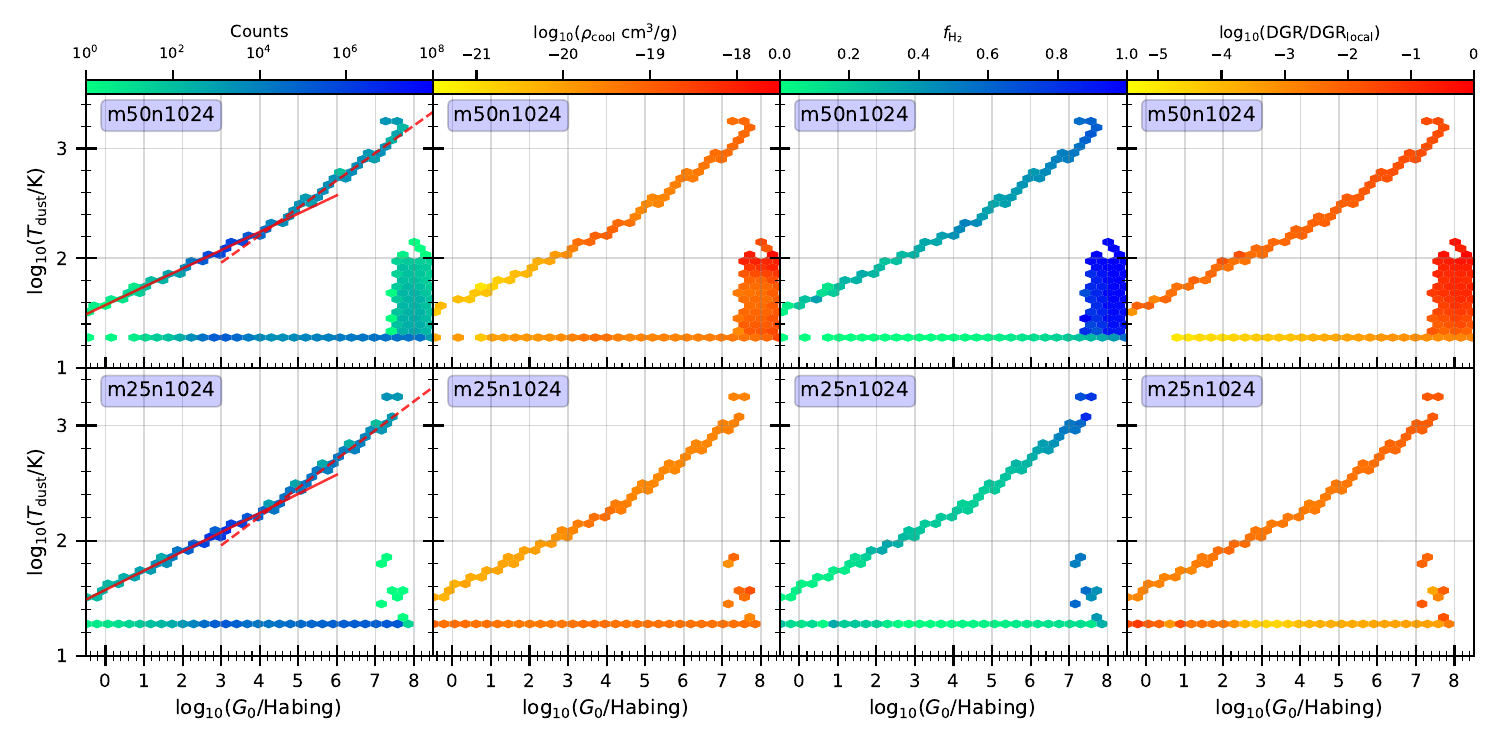}
        \caption{The dust temperature of all galactic particles as a function of the local interstellar radiation field $G_0$ incident upon them, for both our 50 (top row) and 25 (bottom row) Mpc/h boxes. We divide our grid into approximately regular hexagonal bins such that there are 30 along the x-axis, the colour of each quantifying the mean statistic (except for the counts). Each column of the figure places a different property on the colour bar, left-to-right these correspond to: the number of particles, the density of the cool-phase ISM gas, the molecular hydrogen fraction, and the dust-to-gas ratio scaled to that of the local universe \citep{bib:pollack1994}. Furthermore, the first column contains solid and dashed red lines which show the dust temperature scaling as $\sim G_0^{1/6}$ and $\sim G_0^{1/4}$ respectively; these are useful references for our discussion. We observe a population of high-$G_0$ low-$T_\mathrm{dust}$ particles in the larger box which are largely absent in its smaller counterpart. In both boxes, the majority of particles seem to exist along a `primary' sequence for which the dust temperature is positively correlated with ISRF strength. Additionally, there exists a second sequence present at both resolutions, where the dust temperature remains constant at $\sim 10^{1.3} \approx 18$ K at all values of $G_0$. Comparison between columns reveals that the cool-phase density  and molecular hydrogen fractions are tightly correlated as expected. Of particular note is the region of maximal molecular hydrogen abundance in the bottom-right corner of the 50 Mpc/h panel. This region does not appear in the smaller box, highlighting a significant lack of fully molecular gas as seen in Figure \ref{fig:H2_metDtg}.}
        \label{fig:dustHeatingISRF}
    \end{figure*}

    \begin{figure}
        \centering
        \includegraphics[width=\columnwidth]{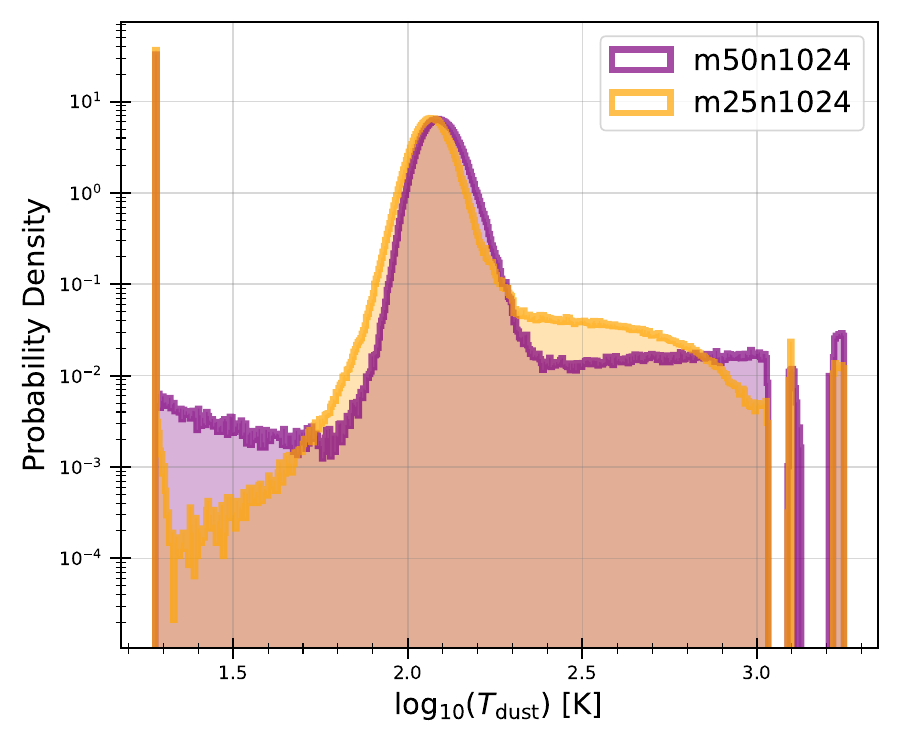}
        \caption{A histogram (500 equal-width bins) showing the dust temperature distribution of all galactic particles in both our 50 Mpc/h (purple) and 25 Mpc/h (orange) simulation boxes, as seen on the y-axis of Figure \ref{fig:dustHeatingISRF}. The first peak at $\sim 10^{1.3} \,\, \mathrm{K} \approx 18 \,\, \mathrm{K}$ corresponds to the constant, low-temperature sequence identified in Figure \ref{fig:dustHeatingISRF}. The abrupt, high-amplitude nature of this peak suggests that the dust is unable to cool any further than this. In both simulations we see a well defined peak at $\sim 10^{2.1} \,\, \mathrm{K} \approx 126 \,\, \mathrm{K}$ which contains significantly more probability than any other region (bar the temperature floor). Both simulations exhibit similar maximal dust temperature values of $\sim 10^{3.25} \,\, \mathrm{K} \approx 1778 \,\, \mathrm{K}$, the 50 Mpc/h box having a slightly larger probability at this value.}
        \label{fig:dustTempHist}
    \end{figure}

    We explore the relationship between the ISRF incident on gas particles in our model and the subsequent temperature of the dust they contain in Figure \ref{fig:dustHeatingISRF}. The top and bottom rows show results from our 50 and 25 Mpc/h boxes respectively, with each column displaying a different physical quantity on the colour bar. For each panel we split the x-axis into 30 (approximately regular) hexagonal bins and calculate the mean statistic within each to display on the colour bar: except in the case of the first column which is a count of the number of particles contained within the bin. Our dataset contains all particles which are identified to exist within a galaxy.

    Across all panels we identify three primary regions of interest: 1) the `primary' sequence where the dust temperature is positively correlated with the ISRF strength; 2) the sequence of constant dust temperature with increasing $G_0$ at $\sim 10^{1.3} \approx 18$ K, the CMB temperature floor; 3) a population of high-$G_0$ low-$T_\mathrm{dust}$ particles which are not at the dust temperature floor, this region is vastly more populated in the 50 Mpc/h simulation.

    Investigating first the primary sequence, we see that in both simulations, the $10^2 \lesssim G_0 / \mathrm{Habing} \lesssim 10^4$ ($10^{1.8} \lesssim T_\mathrm{dust} / \mathrm{K} \lesssim 10^{2.4}$) region contains the largest population of particles. In general, along this sequence we observe that as the ISRF strength increases, so too does the density of the cool-phase ISM and molecular hydrogen fraction. The dust-to-gas ratio does not display a noticeable dependency on $G_0$ in this region compared to that exhibited in others.

    The sequence at which the dust temperature is constant at $\sim 10^{1.3} \mathrm{K} \left(\approx 18 \mathrm{K} \right)$ for the entire range of $G_0$ at first seems counter-intuitive. We would expect that in general, the stronger the radiation heating the dust, the larger the dust temperature will become. However, the efficiency of the radiative cooling scales with the dust temperature and remains comparable in magnitude to the ISRF heating rate. If the dust is sufficiently sparse such that the gas-grain collisions are infrequent, heat exchange from the gas to the dust will be slow, resulting in the constant temperature behaviour we observe. It is also important to note that the temperature of this floor necessarily coincides with the CMB temperature as $T_\mathrm{dust} \geq T_\mathrm{CMB}$, which can be see in Equation \ref{eq:heatBalanceISRF}.
    
    Taking a look at the 50 Mpc/h run, we see that the dust-to-gas ratio increases with $G_0$ along this sequence, until at $G_0 \gtrsim 10^7$ Habing the gas-grain coupling term becomes significant and the temperature of the dust begins to rise. We will discuss in detail the relative strengths of each heating and cooling term (see Equation \ref{eq:heatBalanceISRF}) in Section \ref{sec:results_dustAndISM:heatBalance} next.
    
    The final region to highlight is the population of high-$G_0$ low-$T_\mathrm{dust}$ particles which exist above the temperature-floor. The size of this region varies greatly with resolution, the 50 Mpc/h run containing a much larger population here than the 25 Mpc/h run which has orders of magnitude lower particle counts. Furthermore, whereas the larger box has many populated bins (with counts $> 0$) in this region, its smaller counterpart has only six. Inspecting the position of these bins carefully, we notice that the about half of these six bins do not lie within the same region as the 50 Mpc/h box; instead manifesting at marginally lower $G_0$. Most likely, these particles are those which exist in newly star-forming regions, where the sSFR has recently seen a large increase, but insufficient time has elapsed for this to manifest in the gas temperature. These particles should then soon join the `primary' sequence as their gas is being rapidly heated.
    
    Turning our attention to the physical properties in this region, a striking feature is the abundance of molecular hydrogen. Whilst the smaller box still reaches large abundances towards the high-$G_0$ end of its primary sequence, the larger exhibits a far greater population of fully-molecular gas which is absent in the former. Marrying this contrasting behaviour between resolutions back to our discussion in Section \ref{sec:results_kmtComparison:H2_metDust} (Figure \ref{fig:H2_metDtg} in particular) we can see that the deficiency of particles in this region leads to a lack of fully-molecular gas. As expected, this region also contains large dust-to-gas ratios which are needed to facilitate the production of molecular hydrogen.

    We see a large range of interstellar radiation field strengths within our simulations, varying by $\sim 8.5$ dex from lowest to highest. In contrast, the dust temperature varies by $\sim 2$ dex at most, showing a relatively weak coupling between them. Take for example the 50 Mpc/h box's primary sequence: at $G_0 \approx 1$ Habing, we see that $T_\mathrm{dust} \sim 40$ K. If we increase the ISRF by a factor of 1000, we see that $T_\mathrm{dust} \sim 100$ K -- an increase of only $2.5\times$. This is readily explained if we consider the case where the ISRF is dominant over all other heat sources in the heat balance equation (Equation \ref{eq:heatBalanceISRF}), from which we find $\kappa_\mathrm{dust}T_\mathrm{dust}^4 \sim G_0$. For dust temperatures $< 200$ K, we know that $\kappa_\mathrm{dust} \sim T_\mathrm{dust}^2$, therefore $T_\mathrm{dust} \sim G_0^{1/6}$. When $200 < T_\mathrm{dust} / \mathrm{K} < 1500$ we know that $\kappa_\mathrm{dust} \sim \mathrm{constant}$ and therefore $T_\mathrm{dust} \sim G_0^{1/4}$. We see then that the dust temperature scales more strongly with the ISRF at intermediate temperatures, however this is still a very weak power law, which explains the vast range differences between the ISRF and dust temperatures: for further discussion on how the ISRF seen in our simulations corresponds to the dust temperature please see Appendix \ref{app:Tdust-G0}. Moreover, in the first column of Figure \ref{fig:dustHeatingISRF} we show both the $\sim G_0^{1/6}$ and $\sim G_0^{1/4}$ scaling of the dust temperature in the red solid and dashed lines respectively. This shows clearly that the primary sequence we identified follows these scaling relations closely, with the their boundary occurring at $G_0 \sim 10^{4.3}$ Habing.
    
    In addition to how the dust temperature scales with $G_0$, it is important that we also gain an understanding of the temperature at which the majority of the dust exists, as one of the major changes made in this work was to remove the constant dust reference temperature of 20 K previously used (see Equation \ref{eq:dustGrowthTimescale}). Figure \ref{fig:dustTempHist} shows the probability distribution of particle dust temperatures: purple and orange lines show the 50 and 25 Mpc/h simulation boxes respectively. From this we can recognise key features discussed in Figure \ref{fig:dustHeatingISRF} such as the high-amplitude, abrupt peak at $T_\mathrm{dust} \approx 10^{1.3} \,\, \mathrm{K} \sim 18 \,\, \mathrm{K}$ corresponding to the CMB temperature.
    
    As discussed next in Section \ref{sec:results_dustAndISM:heatBalance}, the ISRF is dominant over all other terms (except in the densest environments where gas-grain collisions are frequent) with regards to heating the dust, and so the dust temperatures found here are largely a direct result of the $G_0$ calculation (see Equation \ref{eq:heatBalanceISRF}). The peak at $T_\mathrm{dust} \approx 10^{2.1} \,\, \mathrm{K} \sim 126 \,\, \mathrm{K}$ corresponds to the highest-count bins in the primary sequence, around which both simulations converge. However, we see that the larger box contains many more particles in the pre-peak region than the smaller box; i.e. more particles with a lower dust temperature. In contrast, the smaller box contains more particles in the post-peak domain, where the dust temperature is greater, than its larger counterpart. Looking back to Figure \ref{fig:dustHeatingISRF}, we can see to which regions these discrepancies correspond. Most obviously, the pre-peak particles are those which sit in the high-$G_0$ low-$T_\mathrm{dust}$ population (most prevalent in the 50 Mpc/h box) and the low-tempeature end of the primary sequence. Particles absent from the high-$G_0$ low-$T_\mathrm{dust}$ population in the 25 Mpc/h box exist instead in the primary sequence, which manifests on the histogram as a peak of larger width. Post-peak we are observing the high-temperature end of the primary sequence, a region in which particles from the smaller box are more abundant until $T_\mathrm{dust} \approx 10^{2.7} \sim 501$ K where their population declines past that of the larger box.

    Comparing the dust temperatures calculated for all particles in our simulations and the constant 20 K dust temperature used previously, we see that fiducial \simba was assuming that all particles had (approximately) the lowest dust temperature possible -- our floor sitting at $\sim 18$ K. Through modelling the local ISRF for each particle and explicitly solving the gas-grain heat balance to calculate the dust temperature, we find dust temperatures two orders of magnitude larger than was previously assumed. Furthermore, we see that particles existing above the temperature floor are likely to posses a dust temperature of $\sim 126$ K, roughly six times the assumed reference value. Order-of-magnitude changes in the dust temperature will subsequently increase the characteristic timescale at which dust grains accrete gas-phase metals (Equation \ref{eq:dustGrowthTimescale}), ultimately decreasing the growth rate. Observations of high-redshift, dusty galaxies have shown the existence of a large population of hot dust \citep{bib:viero2022, bib:jones2023} in the early universe. The modelled polynomial relationship between the dust temperature and redshift reported in \citealt{bib:viero2022} results in an estimate of $T_\mathrm{dust} \sim 74$ K at redshift 6, whilst the redshift $\sim 8.31$ observations from \citealt{bib:jones2023} measure dust temperatures $\gtrsim 90$ K. Whilst not as hot as the temperatures we are commonly seeing in our simulations, these observations are a clear departure from the $\sim 20$ K values thought to be representative; further demonstrating the need for sophisticated dust treatment. Moreover, the sample of observed galaxies are likely not representative of our simulated sample as we are not bound by observational constraints. To offer a like-for-like comparison we would need to think carefully about our target selection in order to mimic that carried out in these works.

    \subsection{Dissecting the Gas-Grain Heat Balance}
    \label{sec:results_dustAndISM:heatBalance}

    \begin{figure}
        \centering
        \includegraphics[width=\columnwidth]{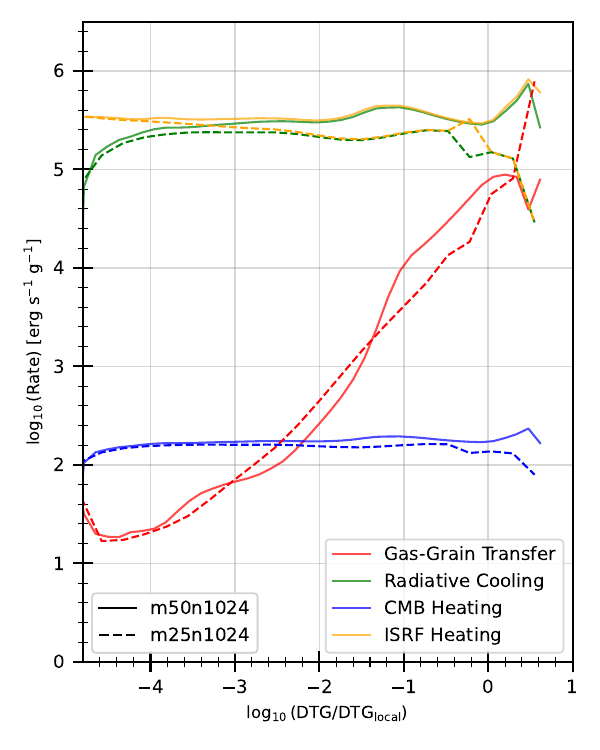}
        \caption{The median dust heating and cooling rate for all galactic particles in both our 50 Mpc/h and 25 Mpc/h runs, displayed by the solid and dashed lines respectively, as a function of the dust-to-gas ratio normalised to the value observed in the local universe \citep{bib:pollack1994}. To compute the median rate we sort the data by its dust-to-gas ratio into 100 regular bins. Each term present in the heat balance equation (see Equation \ref{eq:heatBalanceISRF}) are displayed as follows: the gas-grain heat transfer in red; the radiative cooling in green; the radiative heating from the CMB in blue, and the radiative heating from the ISRF in orange.}
        \label{fig:heatBalance}
    \end{figure}

    The radiative cooling term shown in Figure \ref{fig:heatBalance}, which describes the heat emitted from the grains, is in steady-state thermal equilibrium with the heating terms by construction. In fact, this is how the dust temperature is calculated in Equation \ref{eq:heatBalanceISRF}, and therefore we know that the cooling rate should trace the net heating rate. As the heating is dominated by the ISRF in our case, it should come as no surprise that the cooling rate follows the ISRF heating rate closely. We do see, however, marginal fluctuations between the cooling and ISRF heating terms. It is reasonable to believe that these emerge as a result of minor variations in the physical conditions present. Furthermore, due to the necessity of heat balance, we know that the ISRF (in our case) is responsible for setting the dust temperature. This means that the high dust temperature values present in our simulation could possibly arise from an overestimation of $G_0$.

    As mentioned above, we see that the ISRF is the dominant source of heating across the range of dust abundances present, its heating rate remaining constant at $\sim 10^{5.6} \,\, \mathrm{erg}\,\mathrm{s}^{-1}\,\mathrm{g}^{-1}$ and $\sim 10^{5.4} \,\, \mathrm{erg}\,\mathrm{s}^{-1}\,\mathrm{g}^{-1}$ for the 50 and 25 Mpc/h boxes respectively when $\logten(\mathrm{DTG}/\mathrm{DTG}_\mathrm{local}) < 0$. At high densities, $\logten(\mathrm{DTG}/\mathrm{DTG}_\mathrm{local}) \gtrsim 0$, we see divergent behaviour between our simulations. The smaller box's heating rate falls to $\sim 10^{4.5} \,\, \mathrm{erg}\,\mathrm{s}^{-1}\,\mathrm{g}^{-1}$ by $\logten(\mathrm{DTG}/\mathrm{DTG}_\mathrm{local}) \approx 10^{0.6}$, whilst the larger box's increases to $\sim 10^{5.9} \,\, \mathrm{erg}\,\mathrm{s}^{-1}\,\mathrm{g}^{-1}$. Referring back to Figure \ref{fig:dustHeatingISRF}, we see that the 50 Mpc/h box contains a larger number of particles with a dust-to-gas ratio exceeding that of the local ISM value than its 25 Mpc/h counterpart. Furthermore, the smaller box's most dust abundant particles occur at low-ISRF values; a contrast to their occurrence at high-ISRF values in the larger box. These characteristics explain the discrepant behaviour seen between the boxes at the largest dust abundances. To understand why the ISRF heating rate is generally lower in the 25 Mpc/h box (at all dust abundances) we refer back to Figure \ref{fig:massFunctions}, where it is shown that the larger box contains more particles with intermediate-high SFRs and stellar masses than the smaller. Given that we calculate the ISRF incident on a particle by the sSFR of its neighbouring particles (Equation \ref{eq:localISRF}) it follows that, generally, galactic particles in the 50 Mpc/h box will be irradiated by larger ISRFs than those in the 25 Mpc/h box.

    The rate at which the CMB heats the dust is negligible relative to that of the ISRF. Remaining constant at $\sim 10^{2.2} \,\, \mathrm{erg}\,\mathrm{s}^{-1}\,\mathrm{g}^{-1}$ while $\logten(\mathrm{DTG}/\mathrm{DTG}_\mathrm{local}) \lesssim -2$, there is no appreciable difference between the simulation boxes. However, as $\logten(\mathrm{DTG}/\mathrm{DTG}_\mathrm{local}) > -2$ the 25 Mpc/h box is seen to remain constant whilst the 50 Mpc/h box's heating rate increases marginally. Similar to the behaviour displayed for the ISRF, the smaller box's CMB heating rate drops appreciably at for the most dust-abundant particles whilst the larger box's spikes in this regime. The CMB temperature is constant across all particles -- given that it is inversely proportional to the cosmological scale factor $T_\mathrm{CMB} = T_0\left(1 + Z\right)$ -- and so any inter-particle variance in its heating rate arises from differences in the grain opacity (see Equation \ref{eq:dustOpacity}). Given that its heating rate is $\sim 3$ dex lower than that of the ISRF, it has a negligible contribution to the overall heat balance. However, this does highlight the relevance of the ISRF's inclusion; without which we would find vastly different dust temperatures to those presented in this work.

    The final heating term describes the rate of energy transfer as the gas and grains collide. Unlike the previous terms discussed, this can act to either heat or cool the gas depending on the direction of the temperature gradient. Furthermore, due to the collisional nature of this process, we expect a strong positive correlation with the dust-to-gas ratio, which is clearly seen in Figure \ref{fig:heatBalance}. For example, within dust-sparse gas, $\logten(\mathrm{DTG}/\mathrm{DTG}_\mathrm{local}) < -3$, the collisional heating rate is entirely negligible relative to the other terms, contributing even less than the CMB radiation. However, in dust-rich environments $\logten(\mathrm{DTG}/\mathrm{DTG}_\mathrm{local}) > 0$ the gas-grain heat exchange is significant, reaching rates comparable to that of the ISRF radiation's heating. Slight variations between the two simulated resolutions are observed across the range of dust abundances. The high-resolution run shows a consistent linear scaling, whilst the low-resolution run's median appears more turbulent. Given that collisional processes such as these are highly dependent on environmental specifics, fluctuations in the collisional heating rate between regions of identical dust-to-gas ratio are expected.

\section{Conclusions}
\label{sec:conclusions}

The importance of the link between the ISM and star formation in galaxies is well established. The KMT model has crucially provided a high-fidelity and computationally feasible solution to simulating the co-evolution of metallicity, dust, and molecular hydrogen within galaxy simulations. As our ability to model each of these aspects separately has developed, we now wish to finally connect them so as to lay the groundwork for a fully self-consistent replacement of the KMT model -- as we have attempted to do here. This will allow us to improve each aspect individually as it becomes possible to do so with innovations in modelling methods and computational hardware. The new model presented here combines the non-equilibrium H$_2$ chemistry solver, Grackle, with the dust evolution model of \citet{bib:Li2019}. This allows us to simulate the cosmic evolution of dust and its impact in catalysing the formation of molecular hydrogen, in addition to their joint thermal regulation of the ISM and subsequent influence on star formation. We use this new model, \simbaeor, in two cosmological simulations which we run down to $z \sim 6$. We compare their results with fiducial \simba simulations, in which the cooling, dust evolution, H$_2$ formation and star formation are modelled independently of one-another. Our main findings are summarised below:

\begin{itemize}
    \item We confirm that our simulations are consistent with key \simba results, in addition to various observations, through comparison of the SFR, dust mass and stellar mass functions. \\
    \item Calculating the photometric properties of our galaxies, we find that the brightest observed sources (which have proved difficult to reproduce previously) are reasonably well modelled in our work at redshift 6. We see that the high-redshift dust extinction is larger than expected, which is a result of \simba's artificial feedback suppression. \\
    \item Through post-processing our simulated galaxies to obtain the KMT-estimated H$_2$ fractions, we find that, compared with the results obtained by solving the chemical network, the KMT model overestimates the number of particles with $80\% < f_{\mathrm{H}_2} \leq 100\%$ and underestimates those with $0\% < f_{\mathrm{H}_2}  \leq 20\%$. \\
    \item We find that the ISM of our simulated galaxies are most frequently experiencing incident radiation of $\sim 10^3$ Habing at $z\sim6$; though this varies greatly depending on environmental specifics. Dissecting the gas-grain heat balance, we find that the ISRF dominates all other sources of dust heating in all but the most extreme environments where gas-grain collisions are sufficiently frequent. \\
    \item The expected dust temperature within our model is $\sim126$ K, which is an order-of-magnitude larger than the reference temperature ($20$ K) used in dust model of fiducial \simba. We confirm that this temperature coincides with an ISRF of $\sim 10^3$ Habing.
\end{itemize}

There are many avenues to be explored in continuing this work into the future, the most immediate being investigation of the low-redshift behaviour of our model. Secondly, confirmation/enhancement of the ISRF model is a crucial task in the continuation of this work. Whilst the calculation implemented is well motivated and does not yield unphysical results, the overwhelming significance of the ISRF's contribution to dust heating necessitates further work. The addition of an explicit luminosity-driven calculation (as discussed in \ref{sec:methods:modellingISRF}) is an obvious candidate for such work. Furthermore, the ISRF goes hand-in-hand with the issue of dust shielding, the KMT approximation for which (a scaling with surface metal density) was removed in this work. Essentially, the effect of dust shielding is to attenuate the ISRF heating term in the heat balance equation (Equation \ref{eq:heatBalanceISRF}). Implementing a justified and self-consistent calculation of this attenuation will be important in future works. Finally, the last suggestion we will offer here is a rigorous investigation into the cause of the resolution-dependencies highlighted in this work -- that of the disparate high-$G_0$ low-$T_\mathrm{dust}$ population shown in Figure \ref{fig:dustHeatingISRF} being of particular note. As this population contains the vast majority of fully-molecular and dust-abundant gas particles, it is important that we understand the physical conditions which lead to its existence (or lack thereof); an investigation which was beyond the scope of this work.

\section*{Software}

To analyse our results we used the following astrophysical Python packages: yt \citep{bib:YT} for handling of the raw data contained within simulation snapshots; Caesar \footnote{\url{https://caesar.readthedocs.io/en/latest}} to identify and group particles into galaxies and halos for analysis; pyloser \footnote{\url{https://pyloser.readthedocs.io/en/latest}}, an extension of CAESAR which uses fsps \citep{bib:FSPS_1, bib:FSPS_2} to compute photometry on identified galaxies; and Matplotlib \citep{bib:matplotlib} to create the figures.

\section*{Acknowledgements}

EJ was supported by the STFC. BDS was supported by STFC Consolidated Grant RA5496. DN acknowledges support from the US NSF via Grant AAG-1909153, NASA via grant ATP-80NSSC22K0716 and the Space Telescope Science institution via HST-AR-16145.001. This work used the Cirrus UK National Tier-2 HPC Service at EPCC\footnote{\url{http://www.cirrus.ac.uk}} funded by the University of Edinburgh and EPSRC (EP/P020267/1): access to this machine was given through the Scottish Academic Access grant.
For the purpose of open access, the author has applied a Creative Commons Attribution (CC BY) licence to any Author Accepted Manuscript version arising from this submission.

\section*{Data Availability}

The simulation snapshots, galaxy catalogues and codes from which this work was composed is stored locally, but can be made available to any interested parties upon reasonable request to the authors.



\bibliographystyle{mnras}
\bibliography{thebibliography} 

\begin{thebibliography}{}
\makeatletter
\relax
\def\mn@urlcharsother{\let\do\@makeother \do\$\do\&\do\#\do\^\do\_\do\%\do\~}
\def\mn@doi{\begingroup\mn@urlcharsother \@ifnextchar [ {\mn@doi@} {\mn@doi@[]}}
\def\mn@doi@[#1]#2{\def\@tempa{#1}\ifx\@tempa\@empty \href {http://dx.doi.org/#2} {doi:#2}\else \href {http://dx.doi.org/#2} {#1}\fi \endgroup}
\def\mn@eprint#1#2{\mn@eprint@#1:#2::\@nil}
\def\mn@eprint@arXiv#1{\href {http://arxiv.org/abs/#1} {{\tt arXiv:#1}}}
\def\mn@eprint@dblp#1{\href {http://dblp.uni-trier.de/rec/bibtex/#1.xml} {dblp:#1}}
\def\mn@eprint@#1:#2:#3:#4\@nil{\def\@tempa {#1}\def\@tempb {#2}\def\@tempc {#3}\ifx \@tempc \@empty \let \@tempc \@tempb \let \@tempb \@tempa \fi \ifx \@tempb \@empty \def\@tempb {arXiv}\fi \@ifundefined {mn@eprint@\@tempb}{\@tempb:\@tempc}{\expandafter \expandafter \csname mn@eprint@\@tempb\endcsname \expandafter{\@tempc}}}

\bibitem[\protect\citeauthoryear{Andrews \& Martini}{Andrews \& Martini}{2013}]{bib:andrews2013}
Andrews B.~H.,  Martini P.,  2013, \mn@doi [The Astrophysical Journal] {10.1088/0004-637X/765/2/140}, 765, 140

\bibitem[\protect\citeauthoryear{Aoyama, Hou, Shimizu, Hirashita, Todoroki, Choi  \& Nagamine}{Aoyama et~al.}{2017}]{bib:aoyama2017}
Aoyama S.,  Hou K.-C.,  Shimizu I.,  Hirashita H.,  Todoroki K.,  Choi J.-H.,   Nagamine K.,  2017, \mn@doi [Monthly Notices of the Royal Astronomical Society] {10.1093/mnras/stw3061}, 466, 105

\bibitem[\protect\citeauthoryear{Aoyama, Hou, Hirashita, Nagamine  \& Shimizu}{Aoyama et~al.}{2018}]{bib:aoyama2018}
Aoyama S.,  Hou K.-C.,  Hirashita H.,  Nagamine K.,   Shimizu I.,  2018, \mn@doi [Monthly Notices of the Royal Astronomical Society] {10.48550/ARXIV.1802.04027}

\bibitem[\protect\citeauthoryear{Aoyama, Hirashita  \& Nagamine}{Aoyama et~al.}{2019}]{bib:aoyama2019}
Aoyama S.,  Hirashita H.,   Nagamine K.,  2019, \mn@doi [Monthly Notices of the Royal Astronomical Society] {10.1093/mnras/stz3253}, p. stz3253

\bibitem[\protect\citeauthoryear{Asano, Takeuchi, Hirashita  \& Inoue}{Asano et~al.}{2013}]{bib:Asano2013}
Asano R.~S.,  Takeuchi T.~T.,  Hirashita H.,   Inoue A.~K.,  2013, \mn@doi [Earth, Planets and Space] {10.5047/eps.2012.04.014}, 65, 213

\bibitem[\protect\citeauthoryear{Bekki}{Bekki}{2013}]{bib:bekki2013}
Bekki K.,  2013, \mn@doi [Monthly Notices of the Royal Astronomical Society] {10.1093/mnras/stt589}, 432, 2298

\bibitem[\protect\citeauthoryear{Bekki}{Bekki}{2015}]{bib:bekki2015}
Bekki K.,  2015, \mn@doi [Monthly Notices of the Royal Astronomical Society] {10.1093/mnras/stv165}, 449, 1625

\bibitem[\protect\citeauthoryear{Bigiel, Leroy, Walter, Brinks, De~Blok, Madore  \& Thornley}{Bigiel et~al.}{2008}]{bib:bigiel2008}
Bigiel F.,  Leroy A.,  Walter F.,  Brinks E.,  De~Blok W. J.~G.,  Madore B.,   Thornley M.~D.,  2008, \mn@doi [The Astronomical Journal] {10.1088/0004-6256/136/6/2846}, 136, 2846

\bibitem[\protect\citeauthoryear{Bouwens et~al.,}{Bouwens et~al.}{2015}]{bib:bouwens2015}
Bouwens R.~J.,  et~al., 2015, \mn@doi [The Astrophysical Journal] {10.1088/0004-637X/803/1/34}, 803, 34

\bibitem[\protect\citeauthoryear{Bouwens et~al.,}{Bouwens et~al.}{2021}]{bib:bouwens2021}
Bouwens R.~J.,  et~al., 2021, \mn@doi [The Astronomical Journal] {10.3847/1538-3881/abf83e}, 162, 47

\bibitem[\protect\citeauthoryear{Bromm}{Bromm}{2013}]{bib:bromm2013}
Bromm V.,  2013, \mn@doi [Reports on Progress in Physics] {10.1088/0034-4885/76/11/112901}, 76, 112901

\bibitem[\protect\citeauthoryear{Chiaki \& Wise}{Chiaki \& Wise}{2019}]{bib:chiaki2019}
Chiaki G.,  Wise J.~H.,  2019, \mn@doi [Monthly Notices of the Royal Astronomical Society] {10.1093/mnras/sty2984}, 482, 3933

\bibitem[\protect\citeauthoryear{Choban, Keres, Hopkins, Sandstrom, Hayward  \& Faucher-Giguere}{Choban et~al.}{2022}]{bib:choban2022}
Choban C.~R.,  Keres D.,  Hopkins P.~F.,  Sandstrom K.~M.,  Hayward C.~C.,   Faucher-Giguere C.-A.,  2022, \mn@doi [Monthly Notices of the Royal Astronomical Society] {10.48550/ARXIV.2201.12369}

\bibitem[\protect\citeauthoryear{Choban, Kereš, Sandstrom, Hopkins, Hayward  \& Faucher-Giguère}{Choban et~al.}{2024}]{bib:choban2024}
Choban C.~R.,  Kereš D.,  Sandstrom K.~M.,  Hopkins P.~F.,  Hayward C.~C.,   Faucher-Giguère C.-A.,  2024, A {Dusty} {Locale}: evolution of galactic dust populations from {Milky} {Way} to dwarf-mass galaxies, \mn@doi{10.1093/mnras/stae716}, \url {https://academic.oup.com/mnras/article/529/3/2356/7625602}

\bibitem[\protect\citeauthoryear{{Conroy} \& {Gunn}}{{Conroy} \& {Gunn}}{2010}]{bib:FSPS_2}
{Conroy} C.,  {Gunn} J.~E.,  2010, \mn@doi [\apj] {10.1088/0004-637X/712/2/833}, \href {https://ui.adsabs.harvard.edu/abs/2010ApJ...712..833C} {712, 833}

\bibitem[\protect\citeauthoryear{{Conroy}, {Gunn}  \& {White}}{{Conroy} et~al.}{2009}]{bib:FSPS_1}
{Conroy} C.,  {Gunn} J.~E.,   {White} M.,  2009, \mn@doi [\apj] {10.1088/0004-637X/699/1/486}, \href {https://ui.adsabs.harvard.edu/abs/2009ApJ...699..486C} {699, 486}

\bibitem[\protect\citeauthoryear{Davé, Thompson  \& Hopkins}{Davé et~al.}{2016}]{bib:mufasa2016}
Davé R.,  Thompson R.~J.,   Hopkins P.~F.,  2016, \mn@doi [Monthly Notices of the Royal Astronomical Society] {10.1093/mnras/stw1862}, 462, 3265

\bibitem[\protect\citeauthoryear{Davé, Anglés-Alcázar, Narayanan, Li, Rafieferantsoa  \& Appleby}{Davé et~al.}{2019}]{bib:simba2019}
Davé R.,  Anglés-Alcázar D.,  Narayanan D.,  Li Q.,  Rafieferantsoa M.~H.,   Appleby S.,  2019, \mn@doi [Monthly Notices of the Royal Astronomical Society] {10.1093/mnras/stz937}, 486, 2827

\bibitem[\protect\citeauthoryear{Dayal et~al.,}{Dayal et~al.}{2022}]{bib:dayal2022}
Dayal P.,  et~al., 2022, \mn@doi [Monthly Notices of the Royal Astronomical Society] {10.1093/mnras/stac537}, 512, 989

\bibitem[\protect\citeauthoryear{De~Vis et~al.,}{De~Vis et~al.}{2019}]{bib:deVis2019}
De~Vis P.,  et~al., 2019, \mn@doi [Astronomy \& Astrophysics] {10.1051/0004-6361/201834444}, 623, A5

\bibitem[\protect\citeauthoryear{Donnan et~al.,}{Donnan et~al.}{2022}]{bib:donnan2022}
Donnan C.~T.,  et~al., 2022, \mn@doi [Monthly Notices of the Royal Astronomical Society] {10.1093/mnras/stac3472}, 518, 6011

\bibitem[\protect\citeauthoryear{Dopcke, Glover, Clark  \& Klessen}{Dopcke et~al.}{2011}]{bib:dopcke2011}
Dopcke G.,  Glover S. C.~O.,  Clark P.~C.,   Klessen R.~S.,  2011, \mn@doi [The Astrophysical Journal] {10.1088/2041-8205/729/1/L3}, 729, L3

\bibitem[\protect\citeauthoryear{Draine \& Salpeter}{Draine \& Salpeter}{1979}]{bib:draine1979}
Draine B.~T.,  Salpeter E.~E.,  1979, \mn@doi [The Astrophysical Journal] {10.1086/157165}, 231, 77

\bibitem[\protect\citeauthoryear{Duncan et~al.,}{Duncan et~al.}{2014}]{bib:duncan2014}
Duncan K.,  et~al., 2014, \mn@doi [Monthly Notices of the Royal Astronomical Society] {10.1093/mnras/stu1622}, 444, 2960

\bibitem[\protect\citeauthoryear{Dwek}{Dwek}{1998}]{bib:Dwek1998}
Dwek E.,  1998, \mn@doi [The Astrophysical Journal] {10.1086/305829}, 501, 643

\bibitem[\protect\citeauthoryear{Ferland et~al.,}{Ferland et~al.}{2013}]{bib:ferland2013}
Ferland G.~J.,  et~al., 2013, The 2013 {Release} of {Cloudy}, \url {http://arxiv.org/abs/1302.4485}

\bibitem[\protect\citeauthoryear{Finkelstein et~al.,}{Finkelstein et~al.}{2023}]{bib:finkelstein2023}
Finkelstein S.~L.,  et~al., 2023, The {Complete} {CEERS} {Early} {Universe} {Galaxy} {Sample}: {A} {Surprisingly} {Slow} {Evolution} of the {Space} {Density} of {Bright} {Galaxies} at z {\textasciitilde} 8.5-14.5, \url {http://arxiv.org/abs/2311.04279}

\bibitem[\protect\citeauthoryear{Finlator \& Davé}{Finlator \& Davé}{2008}]{bib:finlator2008}
Finlator K.,  Davé R.,  2008, \mn@doi [Monthly Notices of the Royal Astronomical Society] {10.1111/j.1365-2966.2008.12991.x}, 385, 2181

\bibitem[\protect\citeauthoryear{Furtak, Atek, Lehnert, Chevallard  \& Charlot}{Furtak et~al.}{2020}]{bib:furtak2020}
Furtak L.~J.,  Atek H.,  Lehnert M.~D.,  Chevallard J.,   Charlot S.,  2020, \mn@doi [Monthly Notices of the Royal Astronomical Society] {10.1093/mnras/staa3760}, 501, 1568

\bibitem[\protect\citeauthoryear{Galli \& Palla}{Galli \& Palla}{1998}]{bib:galli1998}
Galli D.,  Palla F.,  1998, The {Chemistry} of the {Early} {Universe}, \url {http://arxiv.org/abs/astro-ph/9803315}

\bibitem[\protect\citeauthoryear{Gardner et~al.,}{Gardner et~al.}{2006}]{bib:gardner2006}
Gardner J.~P.,  et~al., 2006, \mn@doi [Space Science Reviews] {10.1007/s11214-006-8315-7}, 123, 485

\bibitem[\protect\citeauthoryear{Gjergo, Granato, Murante, Ragone-Figueroa, Tornatore  \& Borgani}{Gjergo et~al.}{2018}]{bib:gjergo2018}
Gjergo E.,  Granato G.~L.,  Murante G.,  Ragone-Figueroa C.,  Tornatore L.,   Borgani S.,  2018, \mn@doi [Monthly Notices of the Royal Astronomical Society] {10.1093/mnras/sty1564}, 479, 2588

\bibitem[\protect\citeauthoryear{Granato et~al.,}{Granato et~al.}{2021}]{bib:granato2021}
Granato G.~L.,  et~al., 2021, \mn@doi [Monthly Notices of the Royal Astronomical Society] {10.1093/mnras/stab362}, 503, 511

\bibitem[\protect\citeauthoryear{Haardt \& Madau}{Haardt \& Madau}{2012}]{bib:haardt2012}
Haardt F.,  Madau P.,  2012, \mn@doi [The Astrophysical Journal] {10.1088/0004-637X/746/2/125}, 746, 125

\bibitem[\protect\citeauthoryear{{Habing}}{{Habing}}{1968}]{bib:habing1968}
{Habing} H.~J.,  1968, \bain, \href {https://ui.adsabs.harvard.edu/abs/1968BAN....19..421H} {19, 421}

\bibitem[\protect\citeauthoryear{Harikane et~al.,}{Harikane et~al.}{2022}]{bib:harikane2022}
Harikane Y.,  et~al., 2022, \mn@doi [The Astrophysical Journal Supplement Series] {10.3847/1538-4365/ac3dfc}, 259, 20

\bibitem[\protect\citeauthoryear{Harikane et~al.,}{Harikane et~al.}{2023a}]{bib:harikane2023}
Harikane Y.,  et~al., 2023a, \mn@doi [The Astrophysical Journal Supplement Series] {10.3847/1538-4365/acaaa9}, 265, 5

\bibitem[\protect\citeauthoryear{Harikane et~al.,}{Harikane et~al.}{2023b}]{bib:harikane2023_sfrd}
Harikane Y.,  et~al., 2023b, \mn@doi [The Astrophysical Journal Supplement Series] {10.3847/1538-4365/acaaa9}, 265, 5

\bibitem[\protect\citeauthoryear{Harvey et~al.,}{Harvey et~al.}{2024}]{bib:harvey2024}
Harvey T.,  et~al., 2024, {EPOCHS} {IV}: {SED} {Modelling} {Assumptions} and their impact on the {Stellar} {Mass} {Function} at 6.5 \&lt; z \&lt; 13.5 using {PEARLS} and public {JWST} observations, \mn@doi{10.48550/ARXIV.2403.03908}, \url {https://arxiv.org/abs/2403.03908}

\bibitem[\protect\citeauthoryear{Hirashita}{Hirashita}{2000}]{bib:Hirashita2000}
Hirashita H.,  2000, \mn@doi [Publications of the Astronomical Society of Japan] {10.1093/pasj/52.4.585}, 52, 585

\bibitem[\protect\citeauthoryear{Hirashita \& Kuo}{Hirashita \& Kuo}{2011}]{bib:hirashita2011}
Hirashita H.,  Kuo T.-M.,  2011, \mn@doi [Monthly Notices of the Royal Astronomical Society] {10.1111/j.1365-2966.2011.19131.x}, 416, 1340

\bibitem[\protect\citeauthoryear{Hollenbach \& McKee}{Hollenbach \& McKee}{1979}]{bib:hollenbach1979}
Hollenbach D.,  McKee C.~F.,  1979, \mn@doi [The Astrophysical Journal Supplement Series] {10.1086/190631}, 41, 555

\bibitem[\protect\citeauthoryear{Hollenbach \& McKee}{Hollenbach \& McKee}{1989}]{bib:hollenbach1989}
Hollenbach D.,  McKee C.~F.,  1989, \mn@doi [The Astrophysical Journal] {10.1086/167595}, 342, 306

\bibitem[\protect\citeauthoryear{Hou, Hirashita  \& Michałowski}{Hou et~al.}{2016}]{bib:hou2016}
Hou K.-C.,  Hirashita H.,   Michałowski M.~J.,  2016, \mn@doi [Publications of the Astronomical Society of Japan] {10.1093/pasj/psw085}, 68, 94

\bibitem[\protect\citeauthoryear{Hou, Hirashita, Nagamine, Aoyama  \& Shimizu}{Hou et~al.}{2017}]{bib:hou2017}
Hou K.-C.,  Hirashita H.,  Nagamine K.,  Aoyama S.,   Shimizu I.,  2017, \mn@doi [Monthly Notices of the Royal Astronomical Society] {10.1093/mnras/stx877}, 469, 870

\bibitem[\protect\citeauthoryear{Hou, Aoyama, Hirashita, Nagamine  \& Shimizu}{Hou et~al.}{2019}]{bib:hou2019}
Hou K.-C.,  Aoyama S.,  Hirashita H.,  Nagamine K.,   Shimizu I.,  2019, \mn@doi [Monthly Notices of the Royal Astronomical Society] {10.1093/mnras/stz121}, 485, 1727

\bibitem[\protect\citeauthoryear{Hu, Zhukovska, Somerville  \& Naab}{Hu et~al.}{2019}]{bib:Hu2019}
Hu C.-Y.,  Zhukovska S.,  Somerville R.~S.,   Naab T.,  2019, \mn@doi [Monthly Notices of the Royal Astronomical Society] {10.1093/mnras/stz1481}, 487, 3252

\bibitem[\protect\citeauthoryear{Hunter}{Hunter}{2007}]{bib:matplotlib}
Hunter J.~D.,  2007, \mn@doi [Computing in Science \& Engineering] {10.1109/MCSE.2007.55}, 9, 90

\bibitem[\protect\citeauthoryear{Ishigaki, Kawamata, Ouchi, Oguri, Shimasaku  \& Ono}{Ishigaki et~al.}{2018}]{bib:ishigaki2018}
Ishigaki M.,  Kawamata R.,  Ouchi M.,  Oguri M.,  Shimasaku K.,   Ono Y.,  2018, \mn@doi [The Astrophysical Journal] {10.3847/1538-4357/aaa544}, 854, 73

\bibitem[\protect\citeauthoryear{Jones \& Nuth}{Jones \& Nuth}{2011}]{bib:jones2011}
Jones A.~P.,  Nuth J.~A.,  2011, \mn@doi [Astronomy \& Astrophysics] {10.1051/0004-6361/201014440}, 530, A44

\bibitem[\protect\citeauthoryear{Jones, Witstok, Concas  \& Laporte}{Jones et~al.}{2023}]{bib:jones2023}
Jones G.~C.,  Witstok J.,  Concas A.,   Laporte N.,  2023, New constraints on the molecular gas content of a \$z{\textbackslash}sim8\$ galaxy from {JVLA} {CO}({J}=2-1) observations, \url {http://arxiv.org/abs/2312.05012}

\bibitem[\protect\citeauthoryear{Kannan, Garaldi, Smith, Pakmor, Springel, Vogelsberger  \& Hernquist}{Kannan et~al.}{2022}]{bib:kannan2022}
Kannan R.,  Garaldi E.,  Smith A.,  Pakmor R.,  Springel V.,  Vogelsberger M.,   Hernquist L.,  2022, \mn@doi [Monthly Notices of the Royal Astronomical Society] {10.1093/mnras/stab3710}, 511, 4005

\bibitem[\protect\citeauthoryear{Krumholz}{Krumholz}{2014}]{bib:krumholz2014}
Krumholz M.~R.,  2014, \mn@doi [Monthly Notices of the Royal Astronomical Society] {10.1093/mnras/stt2000}, 437, 1662

\bibitem[\protect\citeauthoryear{Krumholz \& Gnedin}{Krumholz \& Gnedin}{2011}]{bib:KG2011}
Krumholz M.~R.,  Gnedin N.~Y.,  2011, \mn@doi [The Astrophysical Journal] {10.1088/0004-637X/729/1/36}, 729, 36

\bibitem[\protect\citeauthoryear{Krumholz, McKee  \& Tumlinson}{Krumholz et~al.}{2009}]{bib:KMT2009}
Krumholz M.~R.,  McKee C.~F.,   Tumlinson J.,  2009, \mn@doi [The Astrophysical Journal] {10.1088/0004-637X/693/1/216}, 693, 216

\bibitem[\protect\citeauthoryear{Langeroodi et~al.,}{Langeroodi et~al.}{2022}]{bib:langeroodi2022}
Langeroodi D.,  et~al., 2022, Evolution of the {Mass}-{Metallicity} {Relation} from {Redshift} \$z{\textbackslash}approx8\$ to the {Local} {Universe}, \url {http://arxiv.org/abs/2212.02491}

\bibitem[\protect\citeauthoryear{Leitherer et~al.,}{Leitherer et~al.}{1999}]{bib:STARBURST99}
Leitherer C.,  et~al., 1999, \mn@doi [The Astrophysical Journal Supplement Series] {10.1086/313233}, 123, 3

\bibitem[\protect\citeauthoryear{Li, Narayanan  \& Davé}{Li et~al.}{2019}]{bib:Li2019}
Li Q.,  Narayanan D.,   Davé R.,  2019, \mn@doi [Monthly Notices of the Royal Astronomical Society] {10.1093/mnras/stz2684}, 490, 1425

\bibitem[\protect\citeauthoryear{Licquia \& Newman}{Licquia \& Newman}{2015}]{bib:licquia2015}
Licquia T.~C.,  Newman J.~A.,  2015, \mn@doi [The Astrophysical Journal] {10.1088/0004-637X/806/1/96}, 806, 96

\bibitem[\protect\citeauthoryear{Madau \& Dickinson}{Madau \& Dickinson}{2014}]{bib:madau2014}
Madau P.,  Dickinson M.,  2014, \mn@doi [Annual Review of Astronomy and Astrophysics] {10.1146/annurev-astro-081811-125615}, 52, 415

\bibitem[\protect\citeauthoryear{Mattsson}{Mattsson}{2021}]{bib:mattsson2021}
Mattsson L.,  2021, The {Minimal} {Astration} {Hypothesis} -- a {Necessity} for {Solving} the {Dust} {Budget} {Crisis}?, \url {http://arxiv.org/abs/2112.07735}

\bibitem[\protect\citeauthoryear{McKee}{McKee}{1989}]{bib:mckee1989}
McKee C.~F.,  1989, \mn@doi [Symposium - International Astronomical Union] {10.1017/S0074180900125434}, 135, 431

\bibitem[\protect\citeauthoryear{McKinnon, Torrey  \& Vogelsberger}{McKinnon et~al.}{2016}]{bib:mckinnon2016}
McKinnon R.,  Torrey P.,   Vogelsberger M.,  2016, \mn@doi [Monthly Notices of the Royal Astronomical Society] {10.1093/mnras/stw253}, 457, 3775

\bibitem[\protect\citeauthoryear{McKinnon, Torrey, Vogelsberger, Hayward  \& Marinacci}{McKinnon et~al.}{2017}]{bib:mckinnon2017}
McKinnon R.,  Torrey P.,  Vogelsberger M.,  Hayward C.~C.,   Marinacci F.,  2017, \mn@doi [Monthly Notices of the Royal Astronomical Society] {10.1093/mnras/stx467}, 468, 1505

\bibitem[\protect\citeauthoryear{McKinnon, Vogelsberger, Torrey, Marinacci  \& Kannan}{McKinnon et~al.}{2018}]{bib:mckinnon2018}
McKinnon R.,  Vogelsberger M.,  Torrey P.,  Marinacci F.,   Kannan R.,  2018, Simulating galactic dust grain evolution on a moving mesh, \mn@doi{10.48550/ARXIV.1805.04521}, \url {https://arxiv.org/abs/1805.04521}

\bibitem[\protect\citeauthoryear{McLeod et~al.,}{McLeod et~al.}{2023}]{bib:mcleod2023}
McLeod D.~J.,  et~al., 2023, The galaxy {UV} luminosity function at \${\textbackslash}mathbf\{z {\textbackslash}simeq 11\}\$ from a suite of public {JWST} {ERS}, {ERO} and {Cycle}-1 programs, \url {http://arxiv.org/abs/2304.14469}

\bibitem[\protect\citeauthoryear{Navarro-Carrera, Rinaldi, Caputi, Iani, Kokorev  \& van Mierlo}{Navarro-Carrera et~al.}{2023}]{bib:navarroCarrera2023}
Navarro-Carrera R.,  Rinaldi P.,  Caputi K.~I.,  Iani E.,  Kokorev V.,   van Mierlo S.~E.,  2023, Constraints on the {Faint} {End} of the {Galaxy} {Stellar} {Mass} {Function} at z {\textasciitilde} 4-8 from {Deep} {JWST} {Data}, \url {http://arxiv.org/abs/2305.16141}

\bibitem[\protect\citeauthoryear{Olsen, Greve, Narayanan, Thompson, Davé, Rios  \& Stawinski}{Olsen et~al.}{2017}]{bib:sigame2017}
Olsen K.~P.,  Greve T.~R.,  Narayanan D.,  Thompson R.,  Davé R.,  Rios L.~N.,   Stawinski S.,  2017, \mn@doi [The Astrophysical Journal] {10.3847/1538-4357/aa86b4}, 846, 105

\bibitem[\protect\citeauthoryear{{Pollack}, {Hollenbach}, {Beckwith}, {Simonelli}, {Roush}  \& {Fong}}{{Pollack} et~al.}{1994}]{bib:pollack1994}
{Pollack} J.~B.,  {Hollenbach} D.,  {Beckwith} S.,  {Simonelli} D.~P.,  {Roush} T.,   {Fong} W.,  1994, \mn@doi [\apj] {10.1086/173677}, \href {https://ui.adsabs.harvard.edu/abs/1994ApJ...421..615P} {421, 615}

\bibitem[\protect\citeauthoryear{Polzin, Kravtsov, Semenov  \& Gnedin}{Polzin et~al.}{2024}]{bib:polzin2024}
Polzin A.,  Kravtsov A.~V.,  Semenov V.~A.,   Gnedin N.~Y.,  2024, \mn@doi [The Astrophysical Journal] {10.3847/1538-4357/ad32cb}, 966, 172

\bibitem[\protect\citeauthoryear{Priestley, De~Looze  \& Barlow}{Priestley et~al.}{2021}]{bib:priestley2021}
Priestley F.~D.,  De~Looze I.,   Barlow M.~J.,  2021, \mn@doi [Monthly Notices of the Royal Astronomical Society: Letters] {10.1093/mnrasl/slab114}, 509, L6

\bibitem[\protect\citeauthoryear{Rahmati, Pawlik, Raičevi\`{c}  \& Schaye}{Rahmati et~al.}{2013}]{bib:rahmati2013}
Rahmati A.,  Pawlik A.~H.,  Raičevi\`{c} M.,   Schaye J.,  2013, \mn@doi [Monthly Notices of the Royal Astronomical Society] {10.1093/mnras/stt066}, 430, 2427

\bibitem[\protect\citeauthoryear{Relano et~al.,}{Relano et~al.}{2022}]{bib:relano2022}
Relano M.,  et~al., 2022, \mn@doi [Monthly Notices of the Royal Astronomical Society] {10.1093/mnras/stac2108}, 515, 5306

\bibitem[\protect\citeauthoryear{Relaño, Lisenfeld, Hou, De~Looze, Vílchez  \& Kennicutt}{Relaño et~al.}{2020}]{bib:relano2020}
Relaño M.,  Lisenfeld U.,  Hou K.-C.,  De~Looze I.,  Vílchez J.~M.,   Kennicutt R.~C.,  2020, \mn@doi [Astronomy \& Astrophysics] {10.1051/0004-6361/201937087}, 636, A18

\bibitem[\protect\citeauthoryear{Robertson \& Kravtsov}{Robertson \& Kravtsov}{2008}]{bib:robertson2008}
Robertson B.,  Kravtsov A.,  2008, \mn@doi [The Astrophysical Journal] {10.1086/587796}, 680, 1083

\bibitem[\protect\citeauthoryear{Romano, Nagamine  \& Hirashita}{Romano et~al.}{2022}]{bib:romano2022}
Romano L. E.~C.,  Nagamine K.,   Hirashita H.,  2022, \mn@doi [Monthly Notices of the Royal Astronomical Society] {10.1093/mnras/stac1386}, 514, 1461

\bibitem[\protect\citeauthoryear{Schneider, Omukai, Inoue  \& Ferrara}{Schneider et~al.}{2006}]{bib:schneider2006}
Schneider R.,  Omukai K.,  Inoue A.~K.,   Ferrara A.,  2006, \mn@doi [Monthly Notices of the Royal Astronomical Society] {10.1111/j.1365-2966.2006.10391.x}, 369, 1437

\bibitem[\protect\citeauthoryear{Slavin, Dwek  \& Jones}{Slavin et~al.}{2015}]{bib:slavin2015}
Slavin J.~D.,  Dwek E.,   Jones A.~P.,  2015, \mn@doi [The Astrophysical Journal] {10.1088/0004-637X/803/1/7}, 803, 7

\bibitem[\protect\citeauthoryear{Smit, Bouwens, Franx, Illingworth, Labbé, Oesch  \& Van~Dokkum}{Smit et~al.}{2012}]{bib:smit2012}
Smit R.,  Bouwens R.~J.,  Franx M.,  Illingworth G.~D.,  Labbé I.,  Oesch P.~A.,   Van~Dokkum P.~G.,  2012, \mn@doi [The Astrophysical Journal] {10.1088/0004-637X/756/1/14}, 756, 14

\bibitem[\protect\citeauthoryear{Smith et~al.,}{Smith et~al.}{2017}]{bib:grackle2017}
Smith B.~D.,  et~al., 2017, \mn@doi [Monthly Notices of the Royal Astronomical Society] {10.1093/mnras/stw3291}, 466, 2217

\bibitem[\protect\citeauthoryear{Spitzer}{Spitzer}{2004}]{bib:spitzer2004}
Spitzer L.,  2004, Physical processes in the interstellar medium.
Physics textbook, Wiley, Weinheim

\bibitem[\protect\citeauthoryear{Springel \& Hernquist}{Springel \& Hernquist}{2003}]{bib:springel2003}
Springel V.,  Hernquist L.,  2003, \mn@doi [Monthly Notices of the Royal Astronomical Society] {10.1046/j.1365-8711.2003.06206.x}, 339, 289

\bibitem[\protect\citeauthoryear{Stefanon, Bouwens, Labbé, Illingworth, Gonzalez  \& Oesch}{Stefanon et~al.}{2021}]{bib:stefanon2021}
Stefanon M.,  Bouwens R.~J.,  Labbé I.,  Illingworth G.~D.,  Gonzalez V.,   Oesch P.~A.,  2021, \mn@doi [The Astrophysical Journal] {10.3847/1538-4357/ac1bb6}, 922, 29

\bibitem[\protect\citeauthoryear{Tegmark, Silk, Rees, Blanchard, Abel  \& Palla}{Tegmark et~al.}{1997}]{bib:tegmark1997}
Tegmark M.,  Silk J.,  Rees M.,  Blanchard A.,  Abel T.,   Palla F.,  1997, \mn@doi [The Astrophysical Journal] {10.1086/303434}, 474, 1

\bibitem[\protect\citeauthoryear{Tielens \& Hollenbach}{Tielens \& Hollenbach}{1985}]{bib:tielens1985}
Tielens A. G. G.~M.,  Hollenbach D.,  1985, \mn@doi [The Astrophysical Journal] {10.1086/163111}, 291, 722

\bibitem[\protect\citeauthoryear{Tielens, McKee, Seab  \& Hollenbach}{Tielens et~al.}{1994}]{bib:tielens1994}
Tielens A. G. G.~M.,  McKee C.~F.,  Seab C.~G.,   Hollenbach D.~J.,  1994, \mn@doi [The Astrophysical Journal] {10.1086/174488}, 431, 321

\bibitem[\protect\citeauthoryear{Tsai \& Mathews}{Tsai \& Mathews}{1995}]{bib:tsai1995}
Tsai J.~C.,  Mathews W.~G.,  1995, \mn@doi [The Astrophysical Journal] {10.1086/175943}, 448, 84

\bibitem[\protect\citeauthoryear{{Turk}, {Smith}, {Oishi}, {Skory}, {Skillman}, {Abel}  \& {Norman}}{{Turk} et~al.}{2011a}]{bib:YT}
{Turk} M.~J.,  {Smith} B.~D.,  {Oishi} J.~S.,  {Skory} S.,  {Skillman} S.~W.,  {Abel} T.,   {Norman} M.~L.,  2011a, \mn@doi [The Astrophysical Journal Supplement Series] {10.1088/0067-0049/192/1/9}, \href {https://ui.adsabs.harvard.edu/abs/2011ApJS..192....9T} {192, 9}

\bibitem[\protect\citeauthoryear{Turk, Clark, Glover, Greif, Abel, Klessen  \& Bromm}{Turk et~al.}{2011b}]{bib:turk2011}
Turk M.~J.,  Clark P.,  Glover S. C.~O.,  Greif T.~H.,  Abel T.,  Klessen R.,   Bromm V.,  2011b, \mn@doi [The Astrophysical Journal] {10.1088/0004-637X/726/1/55}, 726, 55

\bibitem[\protect\citeauthoryear{Viero, Sun, Chung, Moncelsi  \& Condon}{Viero et~al.}{2022}]{bib:viero2022}
Viero M.~P.,  Sun G.,  Chung D.~T.,  Moncelsi L.,   Condon S.~S.,  2022, \mn@doi [Monthly Notices of the Royal Astronomical Society: Letters] {10.1093/mnrasl/slac075}, 516, L30

\bibitem[\protect\citeauthoryear{Wang et~al.,}{Wang et~al.}{2024}]{bib:wang2024}
Wang T.,  et~al., 2024, The true number density of massive galaxies in the early {Universe} revealed by {JWST}/{MIRI}, \mn@doi{10.48550/ARXIV.2403.02399}, \url {https://arxiv.org/abs/2403.02399}

\bibitem[\protect\citeauthoryear{Weibel et~al.,}{Weibel et~al.}{2024}]{bib:weibel2024}
Weibel A.,  et~al., 2024, Galaxy {Build}-up in the first 1.5 {Gyr} of {Cosmic} {History}: {Insights} from the {Stellar} {Mass} {Function} at \$z{\textbackslash}sim4-9\$ from {JWST} {NIRCam} {Observations}, \mn@doi{10.48550/ARXIV.2403.08872}, \url {https://arxiv.org/abs/2403.08872}

\bibitem[\protect\citeauthoryear{Yates, Kauffmann  \& Guo}{Yates et~al.}{2012}]{bib:yates2012}
Yates R.~M.,  Kauffmann G.,   Guo Q.,  2012, \mn@doi [Monthly Notices of the Royal Astronomical Society] {10.1111/j.1365-2966.2012.20595.x}, 422, 215

\bibitem[\protect\citeauthoryear{Zhukovska, Gail  \& Trieloff}{Zhukovska et~al.}{2008}]{bib:zhukovska2008}
Zhukovska S.,  Gail H.-P.,   Trieloff M.,  2008, \mn@doi [Astronomy \& Astrophysics] {10.1051/0004-6361:20077789}, 479, 453

\makeatother
\end{thebibliography}


\appendix
\section{The Dust Accretion Timescale and Sticking Potential}
\label{app:dustGrowth}

Consider the following situation, spherical dust grains of constant mass density grow through the accretion of gas-phase metals on their surface. The collisional cross-sections of these grains are equal to their geometrical cross-sections, ie. we do not consider electrostatic forces. The composition of the grain determines which metal it must accrete in order to grow, this metal is known as the `key species`, denoted by $\mathrm{X}$ in our equations \citep{bib:zhukovska2008}. When a gas-phase metal strikes a grain, there is a probability, $\xi(T_\mathrm{gr}, T_\mathrm{X})$, that it remains bound to the surface, we call this the `sticking potential'.

With these assumptions in mind, we can construct the rate at which gas-phase metals strike the grain surface as,

\begin{align}
    \label{eq:accretionCollisionRate}
    R_\mathrm{coll} = \pi a^2 \braket{v_\mathrm{X}} n_\mathrm{X}
\end{align}

where $a$ is the radius of the grain, $\braket{v_\mathrm{X}}$ is the average speed of the key gas-phase metals, and $n_\mathrm{X}$ is their corresponding number density \citep{bib:hirashita2011, bib:Asano2013}. We can then write the rate at which the gaseous metal population is depleted through accretion as,

\begin{align}
    \label{eq:accretionMassGrowthRate}
    \frac{dm_\mathrm{acc}}{dt} = R_\mathrm{coll} m_\mathrm{X} \xi(T_\mathrm{gr},T_\mathrm{X})
\end{align}

where $m_\mathrm{X}$ is the atomic mass of the key metal species. As we do not model different grain compositions in our simulation, we simply assume that the accreted species are split equally among the metals present. This allows us to write $n_\mathrm{X}m_\mathrm{X} = \rho_\mathrm{g}Z_\mathrm{g}$ where $\rho_\mathrm{g}$ is the mass density of the gas and $Z_\mathrm{g}$ is the gas metallicity. Furthermore, we assume that the metals are in thermal equilibrium with the gas. Combining equations \ref{eq:accretionCollisionRate} and \ref{eq:accretionMassGrowthRate} with these assumptions we obtain,

\begin{align}
    \label{eq:accretionMassGrowthRate_final}
    \frac{dm_\mathrm{acc}}{dt} \sim \pi a^2 T_\mathrm{g}^{1/2} \rho_\mathrm{g} Z_\mathrm{g} \xi(T_\mathrm{gr},T_\mathrm{g})
\end{align}

where we set $\braket{v_\mathrm{g}} \sim T_\mathrm{g}^{1/2}$ assuming the gas follows a Maxwell-Boltzmann distribution.

We now turn our attention to the rate at which the grains themselves grow through accretion. The mass of a spherical grain with constant mass density, $\sigma$, is simply given by $m_\textrm{gr} = 4/3\pi a^3 \sigma$. Differentiating this with respect to time we find,

\begin{align}
    \label{eq:grainGrowthRate}
    \frac{dm_\mathrm{gr}}{dt} = 4\pi a^2 \dot{a} \sigma
\end{align}

where $\dot{a} \equiv da/dt$. Assuming the grains only grow through accretion, we know that the rate at which metals are depleted from the gas due to accretion must equal the growth rate of the grain. This means that we can equate equations \ref{eq:accretionMassGrowthRate_final} and \ref{eq:grainGrowthRate}, then rearrange to find,

\begin{align}
    \dot{a} \sim \frac{1}{4\sigma}\rho_\mathrm{g}Z_\mathrm{g}T_\mathrm{g}^{1/2} \xi(T_\mathrm{gr},T_\mathrm{g}).
\end{align}

The dust accretion timescale is defined as $\tau_\mathrm{acc} \equiv a/\dot{a}$. Using the above equation we obtain,

\begin{align}
    \label{eq:tauAcc}
    \tau_\mathrm{acc} \simeq \tau_\mathrm{acc,0} \xi^{-1}(T_\mathrm{gr},T_\mathrm{g}) \left( \frac{\rho_\mathrm{ref}}{\rho_\mathrm{g}} \right) \left( \frac{Z_\mathrm{ref}}{Z_\mathrm{g}} \right) \left( \frac{T_\mathrm{ref}}{T_\mathrm{g}} \right)^{1/2}
\end{align}

where we have inserted three characteristic reference values to determine the specific normalisation of the density, metallicity and temperature individually. Furthermore, we have added an overall normalisation to the accretion rate, $\tau_\mathrm{acc,0}$. The three reference values should be chosen such that they represent typical values in the physical scenarios being modelled. The accretion rate normalisation should be set to a reasonable value and tuned to suit the specific use case.

We do this parameterisation to leave us with an equation akin to that prevalent in the literature \citep{bib:Li2019, bib:Asano2013}. Note that as we do not model a grain size distribution, the radius of the grain is absorbed into this normalisation. However, we have not assumed that the sticking potential is unity, as is typically done in the literature. This is because it depends on the grain temperature, a parameter of key focus in our work.

As discussed in Section \ref{sec:methods:dustGrowthAndDestruction:metalAccretion}, the exact form of the sticking potential is incredibly complex, so much so that it is impracticable to include in our current simulations which do not track grain compositions or the grain size distribution, in addition to the chemistry or cooling required for grains of specific compositions. We can reason however, that in general, the sticking potential must depend on the energy of the metal-grain collision. If they collide violently, at high energies, then we expect there to be a low probability that the metal will remain bound to the grain surface. Oppositely, if the collision takes place at low energies, the metal will have a high chance of remaining bound to the grain.

The hydrogen-grain sticking potential presented in \citealt{bib:hollenbach1979} is a good example of such behaviour,

\begin{align}
    \label{eq:stickingPotential}
    \xi(T_\mathrm{gr}, T_\mathrm{g}) \sim \frac{1}{\sqrt{T_\mathrm{gr} + T_\mathrm{g}} + 1}
\end{align}

where we have neglected to include the higher-power terms of $T_\mathrm{g}$ presented in the original paper for simplicity. We see that this equation behaves as reasoned above -- as the total temperature approaches zero, the sticking potential approaches unity, ie. there is a 100\% chance that the metal remains bound to the grain surface. Oppositely, as the temperature of one or both of the particles approaches infinity, the sticking potential approaches zero, ie. the metal will never remain bound to the grain surface. 

However, the issue with implementing this equation into our model is its dependence on the gas temperature, which would affect the accretion timescale's behaviour significantly. A complete inclusion of the sticking potential, as a function of both the gas and grain temperatures, would require the development and implementation of an entirely different model designed for this purpose. We therefore make the choice to adopt a sticking potential of the form,

\begin{align}
    \xi(T_\mathrm{gr}) \sim T_\mathrm{gr}^{-1/2}\,.
\end{align}

which introduces only a weak scaling with the dust temperature, ensuring the continued validity of the model with these changes. We can then re-write the accretion timescale as,

\begin{align}
    \tau_\mathrm{acc} \simeq \tau_\mathrm{ref} \left( \frac{\rho_\mathrm{ref}}{\rho_\mathrm{g}} \right) \left( \frac{Z_\mathrm{ref}}{Z_\mathrm{g}} \right) \left( \frac{T_\mathrm{gr}}{T_\mathrm{g}} \right)^{1/2}
\end{align}

where $\tau_\mathrm{ref}$ is a constant; please see Section \ref{sec:methods:dustGrowthAndDestruction:metalAccretion} for details on the reference values used in our work. Comparing this with equation \ref{eq:tauAcc} you will see that in total we have essentially set $T_\mathrm{ref} = T_\mathrm{gr}$; this is the claim we make in Section \ref{sec:methods:dustGrowthAndDestruction:metalAccretion}.

In conclusion, whilst we want the dust temperature to appear in the accretion rate equation (as it makes intuitive sense for this to be the case and provides better self-consistency between models) we carry this out in the simplest way possible. We have made many simplifying assumptions in lieu of fully modelling the grain sizes and compositions, and taken advantage of the fact that the model of \citealt{bib:Li2019} is sub-grid with four free-parameters which require tuning. A result of this inherent ambiguity is that we can add a simple dependence on the dust temperature, which acts to increase/decrease the rate of accretion subtly, whilst allowing the rest of our dust model to feed back to the rate at which it grows. To ensure the validity of our model after these changes, we re-tuned the $\tau_\mathrm{acc,0}$ parameter (renaming it $\tau_\mathrm{ref}$ for clarity) such that the redshift $\sim 6$ stellar mass function it predicts is in alignment with contemporary results from both observation and simulation.

\section{The ISRF's Impact on Dust Temperature}
\label{app:Tdust-G0}

\begin{figure}
    \centering
    \includegraphics[width=1.0\linewidth]{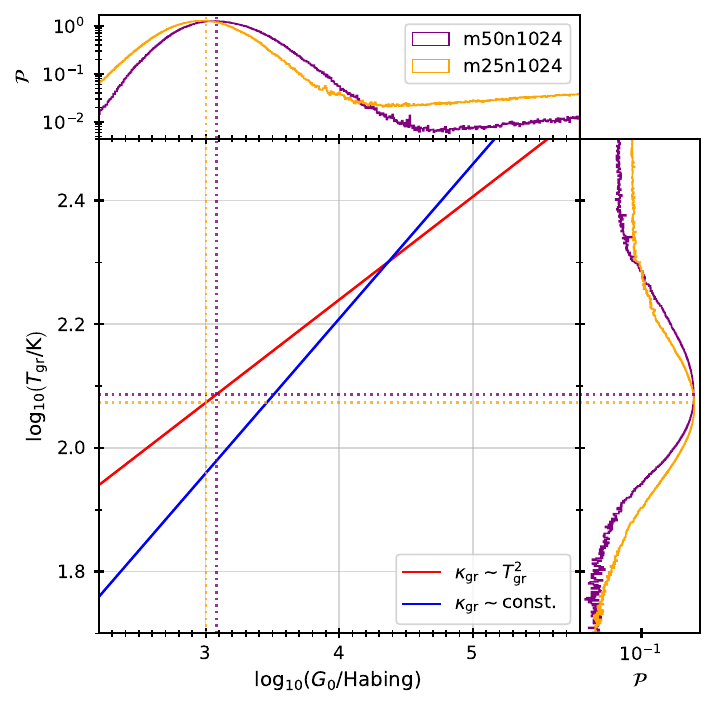}
    \caption{The relationship between the ISRF strength and dust temperature as modelled in our simulation. On the primary axis we show how the dust temperature scales with $G_0$ assuming that the ISRF is the dominant source of heating. We show this for two grain opacities, one which scales quadratically with the dust temperature (red) and one which is constant (blue). The marginal axes show the normalised number density of galactic particles from our 50 Mpc/h and 25 Mpc/h simulations; which we colour purple and orange respectively. Identically coloured dotted construction lines are drawn across both the marginal and primary axes to show the values of peak probability.}
    \label{fig:Tdust-G0}
\end{figure}

Figure \ref{fig:Tdust-G0} shows the relationship between the dust temperature and $G_0$ in the ISRF-dominated regime (see Equation \ref{eq:heatBalanceISRF}). We show this for two grain opacities, $\kappa_\mathrm{gr} \sim T_\mathrm{gr}^2$ (red) and $\kappa_\mathrm{gr} \sim \mathrm{constant}$ (blue), which apply at low and moderate dust temperatures respectively. On the marginal x-axis we show the probability density, $\mathcal{P}$, of ISRF strengths within our simulated galaxies; purple representing our 50 Mpc/h box and orange representing our 25 Mpc/h box. The marginal y-axis shows an equivalent histogram for the dust temperature. Furthermore, we add dotted construction lines indicating the position of the peaks within each distribution.

As detailed in Section \ref{sec:results_dustAndISM:ISRF}, we find that our simulated galaxies contain a large proportion of dust at $T_\mathrm{gr} \sim 126$ K -- here we wish to give further explanation to the existence of this population. Figure \ref{fig:heatBalance} shows that the ISRF is dominant over all other dust heating processes, except at extreme densities where gas-grain collisions become significant. Therefore, if we ignore both the CMB and gas-grain heating terms from Equation \ref{eq:heatBalanceISRF} we find that $T_\mathrm{gr}^4 \sim G_0 / \kappa_\mathrm{gr}$, where the dust opacity, $\kappa_\mathrm{gr}$ is a function of the dust temperature.

Ignoring the limit in which grains are sublimated, we model the grain opacity as $\sim T_\mathrm{gr}^2$ when $T_\mathrm{gr} < 200$ K, and $\sim \mathrm{constant}$ when $200 < T_\mathrm{gr} < 1500$ K. For each of these opacities, assuming ISRF domination, we can then calculate the dust temperature resultant from varying ISRF strengths. This relationship is shown on the primary axis in Figure \ref{fig:Tdust-G0}, with the quadratic dust opacity scaling in red, and its constant counterpart in blue.

On the marginal y-axis we show the probability density of galactic dust temperatures from our simulation, and draw dotted construction lines across the axis at the peak value. As stated previously, we see that this peaks occurs at $\sim 10^{2.1} \sim 126$ K. In similar fashion, we also show the probability density of galactic ISRF strengths on the marginal x-axis. Drawing construction lines accordingly we see that the histograms peak at $\sim 10^3$ Habing.

If we now use this typical $G_0$ value and calculate the dust temperature in the regime of ISRF-domination as explained above, we find that, when the grain opacity is scaling quadratically, the dust temperature is $\sim 126$ K. This calculation is illustrated on the plot, as for each simulation the construction lines can be seen to cross on the scaling relation. Ultimately, this analysis confirms that the dust temperatures present in our simulations are a direct result of the ISRF as it is dominant over all other heating terms.


\bsp	
\label{lastpage}
\end{document}